\documentclass[11pt]{article}

\usepackage[margin=1in]{geometry}
\usepackage[T1]{fontenc}
\usepackage[utf8]{inputenc}
\usepackage{amsmath,amssymb,amsfonts,bm}
\usepackage{graphicx}
\usepackage{siunitx}
\usepackage{booktabs}
\usepackage{cite}
\usepackage[hidelinks]{hyperref}
\usepackage{url}

\graphicspath{{./images/}}
\title{A flexible framework for treatment effect inference in longitudinal clinical studies with skewed outcomes}

\author{
Kazushi Maruo$^{1}$, Ryota Ishii$^{1}$, Yusuke Yamaguchi$^{2}$,\\
Toshio Shimokawa$^{3}$, Tomoyuki Sugimoto$^{4}$, and Masahiko Gosho$^{1}$\\[0.75em]
\small $^{1}$Department of Biostatistics, Institute of Medicine, University of Tsukuba, Ibaraki, Japan\\
\small $^{2}$Biostatistics, Data Science, Astellas Pharma Global Development, Inc., Illinois, USA\\
\small $^{3}$Department of Biostatistics, Faculty of Medicine, Wakayama Medical University, Wakayama, Japan\\
\small $^{4}$Division of Mathematical Science, Department of Systems Innovation,\\
\small Graduate School of Engineering Science, University of Osaka, Osaka, Japan\\[0.75em]
\small Correspondence: Kazushi Maruo, \href{mailto:kazushi.maruo@gmail.com}{kazushi.maruo@gmail.com}
}

\date{September 2026}

\begin{document}
\maketitle

\begin{abstract}
Longitudinal continuous outcomes in clinical trials are commonly analyzed using mixed models for repeated measures (MMRM) under normality assumptions. However, many clinical outcomes are skewed, making mean-based treatment effects difficult to interpret and potentially reducing statistical efficiency.
The Box--Cox MMRM (BCMMRM) approach accommodates skewness by enabling inference on model-based median differences via inverse transformation. However, BCMMRM typically assumes a common transformation parameter across treatment groups and time points. When distributional shapes differ between groups or evolve over time, this assumption may lead to biased treatment effect. Furthermore, when treatment affects not only central tendency but also distributional shape or tail behavior, treatment effects may not be adequately characterized by a single location summary such as the median.

We propose the Box--Cox multivariate regression (BCMVR) framework for longitudinal data with skewed outcomes. BCMVR relaxes this restriction by allowing transformation parameters to vary across groups and time points. The framework enables inference based on interpretable summaries, including median differences and a probability-based treatment effect quantifying the probability that a randomly selected patient in one group has a better outcome than one in another group. This measure integrates information over the entire outcome distribution and provides a complementary summary when distributional shapes differ.
Simulation studies demonstrate that BCMMRM can produce biased estimates when distributions differ in shape, whereas BCMVR provides nearly unbiased estimation. The probability-based measure achieves a favorable balance between robustness and statistical efficiency. The proposed framework provides a flexible and interpretable approach to treatment effect inference under distributional heterogeneity.
\end{abstract}

\noindent\textbf{Keywords:} multivariate analysis; missing values; Box--Cox transformation; model misspecification; R package

\medskip
\noindent\textit{Preprint. This manuscript has been submitted for peer review.}

\section{Introduction}
We consider randomized controlled trials in which a continuous outcome is measured repeatedly over time. In longitudinal data, missingness due to participant dropout or other reasons is almost unavoidable. Consequently, even when the primary interest lies in the treatment effect at a single time point, outcomes at other time points often need to be appropriately incorporated into the analysis in order to account for the impact of missing data. Mixed models for repeated measures (MMRM)\cite{mallinckrodt_etal_01} are widely used as a primary analysis method in clinical trials, as they provide valid inference on treatment effects under the assumption that the missing-data mechanism is missing at random (MAR).

MMRM is a standard and useful framework for mean-based inference under MAR. However, MMRM assumes a multivariate normal distribution for the error terms, an assumption that is rarely satisfied in practice. When departures from normality are substantial and the error distribution is highly skewed, concerns arise regarding both the interpretability and the efficiency of treatment effect estimation. In particular, mean differences may be strongly influenced by tail behavior and may not adequately represent treatment effects when the outcome distribution is asymmetric or heavy-tailed. As an alternative approach for skewed outcomes, rank-based nonparametric methods have also been considered. However, there remain relatively few readily implementable approaches that simultaneously accommodate longitudinal outcomes, covariate adjustment, and missing-at-random (MAR) missingness while directly targeting interpretable probability-based treatment effects.

To address this issue, Maruo \textit{et al}. \cite{maruo_etal_17, bcmixed} proposed a Box--Cox MMRM approach, hereafter referred to as BCMMRM, which applies the Box--Cox transformation \cite{box_cox_64} to the response variable and conducts inference on between-group differences in model-based medians via the inverse transformation. This approach has been shown to mitigate limitations of standard MMRM and to improve both interpretability and estimation efficiency for treatment effect estimation. Nevertheless, BCMMRM assumes a common transformation parameter across treatment groups and time points. When the shape of the error distribution varies substantially across groups or over time, this assumption may lead to biased estimation of treatment effects. Furthermore, even when model-based medians are similar between groups, treatment effects may still differ substantially if the underlying distributional shapes or tail behaviors are different. In such settings, comparisons based solely on a single location summary may provide only a limited characterization of treatment effects.

In this study, we extend the BCMMRM framework by allowing the transformation parameters to vary across treatment groups and time points. Under this extension, the unit of the transformed outcome differs by group and time, rendering the standard MMRM framework inapplicable. To accommodate this setting, we propose a new approach based on fitting multivariate regression models separately within each group, termed the multivariate regression model with the Box--Cox transformation (BCMVR). Analogous to BCMMRM, we develop inference procedures for between-group differences in model-based medians and further introduce a novel probability-based measure of treatment effect. This probability-based measure quantifies the likelihood that a randomly selected individual from one group has a better outcome than a randomly selected individual from another group, without relying solely on comparisons of means or medians.

The objective of this study is to develop a BCMVR framework for treatment effect inference in longitudinal clinical trial data with potentially skewed outcome distributions. The proposed framework enables inference based on interpretable distributional summaries of treatment effects, including the median difference and a probability-based measure that reflects the probability that a patient in one treatment group has a better outcome than a patient in another group. The BCMVR approach therefore provides a unified inferential framework that extends conventional MMRM and BCMMRM methodology while accommodating distributional heterogeneity in longitudinal outcomes. To support practical use, we also provide an R package, \texttt{bcmvr}, implementing the proposed methodology.

The remainder of this paper is organized as follows. In Section 2, we present the analysis model and develop inference procedures for the model parameters. In Section 3, we develop inference procedures for treatment effect measures, including differences in model-based medians and probability-based measures. Section 4 evaluates the performance of the proposed method through simulation studies and an illustrative application, and Section 5 concludes with a discussion.
\section{Model Specification and Parameter Inference}
\subsection{Analysis model}
We first specify the analysis model for the longitudinal outcomes.
We consider a longitudinal study with treatment groups indexed by $g=1,\ldots,G$. 
Within group $g$, participants are indexed by $i_g=1,\ldots,N_g$, and measurements are scheduled at discrete time points $t=1,\ldots,T$. 
Let $y_{i_g(j)}$ denote the $j$th observed response for participant $i_g$ $(j=1,\ldots,n_{i_g})$, where $n_{i_g}\le T$ due to missing observations. 
Equivalently, when indexing by nominal time, $y_{i_gt}$ denotes the response of participant $i_g$ at time $t$, if observed. 
When no observations are missing or the missingness is monotone,
the observed-response index $j$ aligns with the nominal time index $t$;
in this case, we write $y_{i_g(t)} = y_{i_gt}$ for the observation at time $t$.
For participant $i_g$, the vector of observed responses is denoted by $\bm{y}_{i_g}=(y_{i_g(1)},\ldots,y_{i_g(n_{i_g})})^{\top}$. 
Let $x_{i_gk}$ denote the $k$th covariate for participant $i_g$, with $k=1,\ldots,K$; the intercept is included by setting $k=0$ and $x_{i_g0}=1$.

The model is specified as
\[
\bm{z}_{i_g}(\bm{\lambda}_{gi_g})
= \bm{\beta}_{gi_g}\bm{x}_{i_g} + \bm{\varepsilon}_{i_g},
\qquad
\bm{\varepsilon}_{i_g} \sim \mathrm{MVN}_{n_{i_g}}(\bm{0}, \Sigma_{gi_g}).
\]
Here, $\mathrm{MVN}_{n_{i_g}}$ denotes the $n_{i_g}$-dimensional multivariate
normal distribution. The same indexing convention applies to the transformed responses $\bm{z}_{i_g}$,
that is, the notations $z_{i_g(j)}$ and $z_{i_gt}$ are interpreted analogously
as $y_{i_g(j)}$ and $y_{i_gt}$, respectively.

For a participant $i_g$ with no missing observations $(n_{i_g}=T)$, we set
$\bm{\lambda}_{gi_g}=\bm{\lambda}_g=(\lambda_{g1},\ldots,\lambda_{gT})^{\top}$ and
$\bm{z}_{i_g}=(z_{i_g1},\ldots,z_{i_gT})^{\top}$, where the transformed response
$z_{i_gt}$ is defined as the Box--Cox transformation of $y_{i_gt}$ with parameter
$\lambda_{gt}$,
\[
z_{i_gt}
=
\begin{cases}
\displaystyle \frac{y_{i_gt}^{\lambda_{gt}} - 1}{\lambda_{gt}},
& \lambda_{gt} \neq 0, \\[6pt]
\log(y_{i_gt}),
& \lambda_{gt} = 0.
\end{cases}
\]

The covariate vector is $\bm{x}_{i_g}=(1,x_{i_g1},\ldots,x_{i_gK})^{\top}$, and the
regression coefficient matrix is specified as
\[
\bm{\beta}_{gi_g}
= \bm{\beta}_g
=
\begin{pmatrix}
\beta_{g10} & \beta_{g11} & \cdots & \beta_{g1K} \\
\beta_{g20} & \beta_{g21} & \cdots & \beta_{g2K} \\
\vdots      & \vdots      & \ddots & \vdots      \\
\beta_{gT0} & \beta_{gT1} & \cdots & \beta_{gTK}
\end{pmatrix}.
\]
The matrix structure of $\bm{\beta}_g$ allows the regression coefficients to vary
across time points. This flexibility is essential because the transformed
responses $z_{i_gt}$ may have different units across time due to the
time-specific Box--Cox transformation parameters $\lambda_{gt}$.

The error vector is defined as
$\bm{\varepsilon}_{i_g}=(\varepsilon_{i_g1},\ldots,\varepsilon_{i_gT})^{\top}$,
and the covariance matrix is denoted by $\Sigma_{gi_g}=\Sigma_g$.
The covariance structure is parameterized by
$\bm{\alpha}_g=(\alpha_{g1},\ldots,\alpha_{gM})^{\top}$, where $\bm{\alpha}_g$
consists of variance and covariance parameters. Under an unstructured covariance
specification (UN), the number of covariance parameters is $M=T(T+1)/2$.

For participants with missing observations $(n_{i_g}<T)$,
$\bm{\lambda}_{gi_g}$, $\bm{\beta}_{gi_g}$, and $\Sigma_{gi_g}$ correspond to the
subvectors or submatrices obtained by removing the elements associated with the
missing time points from $\bm{\lambda}_g$, $\bm{\beta}_g$, and $\Sigma_g$,
respectively.

The parameter vector to be estimated for group $g$ is given by
\[
\bm{\theta}_g
=
\bigl(
\bm{\lambda}_g^{\top},
\ \mathrm{vec}(\bm{\beta}_g)^{\top},
\ \bm{\alpha}_g^{\top}
\bigr)^{\top},
\]
where $\mathrm{vec}(\bm{\beta}_g)$ denotes the vector obtained by stacking the
columns of $\bm{\beta}_g$.

\subsection{Parameter inference}
Parameter estimation is carried out by maximum likelihood using a two-stage
optimization strategy. For a fixed value of the Box--Cox transformation
parameters $\bm{\lambda}_g$, the remaining parameters
$\mathrm{vec}(\bm{\beta}_g)$ and $\bm{\alpha}_g$ (equivalently, $\Sigma_g$)
are updated by solving the likelihood score equations via a
ridge-stabilized Newton--Raphson algorithm.
The transformation parameters $\bm{\lambda}_g$ are then estimated by maximizing
the profile log-likelihood obtained after plugging in the conditional
maximizers $\widehat{\bm{\beta}}_g(\bm{\lambda}_g)$ and
$\widehat{\bm{\alpha}}_g(\bm{\lambda}_g)$, using a quasi-Newton method.
The resulting maximum likelihood estimator for group $g$ is denoted by
$\widehat{\bm{\theta}}_g
=
(\widehat{\bm{\lambda}}_g^{\top},
\ \mathrm{vec}(\widehat{\bm{\beta}}_g)^{\top},
\ \widehat{\bm{\alpha}}_g^{\top})^{\top}$.

Inference for the model parameters is conducted based on the asymptotic
distribution of the maximum likelihood estimator
$\widehat{\bm{\theta}}_g$.
Let $\mathbf{H}_g$ denote the Hessian matrix of the log-likelihood and
$\mathbf{J}_g$ the matrix defined by the outer products of the individual
likelihood derivatives with respect to $\bm{\theta}_g$, that is,
\[
\mathbf{H}_g
=
\frac{\partial^2 \ell_g(\bm{\theta}_g)}
{\partial \bm{\theta}_g \partial \bm{\theta}_g^{\top}},
\qquad
\mathbf{J}_g
=
\sum_{i_g=1}^{N_g}
\left(
\frac{\partial \ell_{i_g}(\bm{\theta}_g)}
{\partial \bm{\theta}_g}
\right)
\left(
\frac{\partial \ell_{i_g}(\bm{\theta}_g)}
{\partial \bm{\theta}_g}
\right)^{\top},
\]
where $\ell_g(\bm{\theta}_g)=\sum_{i_g=1}^{N_g}\ell_{i_g}(\bm{\theta}_g)$ denotes the log-likelihood for group $g$.

In practice, these matrices are evaluated at the maximum likelihood estimator $\widehat{\bm{\theta}}_g$, yielding
$\widehat{\mathbf{H}}_g=\mathbf{H}_g(\widehat{\bm{\theta}}_g)$ and $\widehat{\mathbf{J}}_g=\mathbf{J}_g(\widehat{\bm{\theta}}_g)$.
The model-based variance estimator of $\widehat{\bm{\theta}}_g$ is defined as
\[
\hat{V}^{(\mathrm{M})}_{\bm{\theta}_g}
=
(-\widehat{\mathbf{H}}_g)^{-1},
\]
whereas the robust variance estimator is given by
\[
\hat{V}^{(\mathrm{R})}_{\bm{\theta}_g}
=
(-\widehat{\mathbf{H}}_g)^{-1}
\widehat{\mathbf{J}}_g
(-\widehat{\mathbf{H}}_g)^{-1}.
\]

Although an unstructured covariance model is assumed within each group and the
Box--Cox transformation allows for a wide range of distributional shapes,
the robust variance estimator
$\hat{V}^{(\mathrm{R})}_{\bm{\theta}_g}$
is employed throughout this study to account for potential departures from the
assumed distributional and covariance assumptions.
Explicit expressions of $\mathbf{H}_g$ and $\mathbf{J}_g$ are provided in the
Supporting Information.

\section{Inference for Treatment Effect Measures}
In this section, using the estimated model parameters, we conduct inference for treatment effect measures at specific time points, derived from the fitted
model, to characterize between-group differences in longitudinal outcomes.
\subsection{Median difference}

For group $g$ at time point $t$, the model-based conditional median
$\xi_{gt}=\xi_{gt}(\bm{\theta}_g)$, evaluated at the average covariate profile, is defined as
\[
\xi_{gt}
=
\begin{cases}
\left(
\lambda_{gt}\,\bm{\beta}_{gt}^{\top}\bar{\bm{x}} + 1
\right)^{1/\lambda_{gt}},
&
\lambda_{gt} \neq 0,
\\[8pt]
\exp\!\left(
\bm{\beta}_{gt}^{\top}\bar{\bm{x}}
\right),
&
\lambda_{gt} = 0,
\end{cases}
\]
which is obtained by applying the inverse Box--Cox transformation to the
model-based mean on the transformed scale.
Here, $\bm{\beta}_{gt}=(\beta_{gt0},\ldots,\beta_{gtK})^{\top}$ denotes the
time-specific regression coefficient vector and
$\bar{\bm{x}}=(1,\bar{x}_1,\ldots,\bar{x}_K)^{\top}$ is the covariate vector,
where $\bar{x}_k$ represents the sample mean of covariate $k$ across all participants
pooled across groups.
The plug-in estimator is obtained by replacing $\bm{\theta}_g$ with its maximum
likelihood estimator $\widehat{\bm{\theta}}_g$, yielding
$\widehat{\xi}_{gt}=\xi_{gt}(\widehat{\bm{\theta}}_g)$.

Inference for $\widehat{\xi}_{gt}$ is conducted using the delta method in
conjunction with the robust variance estimator
$\hat{V}^{(\mathrm{R})}_{\bm{\theta}_g}$.
Specifically, the asymptotic variance of $\widehat{\xi}_{gt}$ is approximated by
\[
\widehat{\mathrm{Var}}\!\left(\widehat{\xi}_{gt}\right)
=
\nabla_{\bm{\theta}_g}\xi_{gt}(\widehat{\bm{\theta}}_g)^{\top}
\hat{V}^{(\mathrm{R})}_{\bm{\theta}_g}
\nabla_{\bm{\theta}_g}\xi_{gt}(\widehat{\bm{\theta}}_g),
\]
where $\nabla_{\bm{\theta}_g}\xi_{gt}(\widehat{\bm{\theta}}_g)$ denotes the gradient
of $\xi_{gt}(\bm{\theta}_g)$ with respect to $\bm{\theta}_g$, evaluated at
$\bm{\theta}_g=\widehat{\bm{\theta}}_g$. Details on the derivation and explicit expressions of the gradient vector
$\nabla_{\bm{\theta}_g}\xi_{gt}(\bm{\theta}_g)$ used in the delta method are
provided in the Supporting Information.

The median difference between groups $g_1$ and $g_2$ at time point $t$ is defined as
\[
\Delta^{(m)}_{g_1 g_2 t}
=
\xi_{g_1 t}-\xi_{g_2 t},
\]
with the corresponding plug-in estimator
\[
\widehat{\Delta}^{(m)}_{g_1 g_2 t}
=
\widehat{\xi}_{g_1 t}-\widehat{\xi}_{g_2 t}.
\]
Assuming independent samples across groups, the variance estimator is given by
\[
\widehat{\mathrm{Var}}\!\left\{\widehat{\Delta}^{(m)}_{g_1 g_2 t}\right\}
=
\widehat{\mathrm{Var}}\!\left(\widehat{\xi}_{g_1 t}\right)
+
\widehat{\mathrm{Var}}\!\left(\widehat{\xi}_{g_2 t}\right).
\]

A Wald-type $100(1-\alpha)\%$ confidence interval for $\Delta^{(m)}_{g_1 g_2 t}$ is given by
\[
\widehat{\Delta}^{(m)}_{g_1 g_2 t}
\ \pm\
z_{1-\alpha/2}
\sqrt{
\widehat{\mathrm{Var}}\!\left\{\widehat{\Delta}^{(m)}_{g_1 g_2 t}\right\}
},
\]
where $z_{1-\alpha/2}$ denotes the $(1-\alpha/2)$ quantile of the standard normal
distribution.
To test the null hypothesis $H_0:\Delta^{(m)}_{g_1 g_2 t}=0$, we use the Wald
test statistic
\[
Z^{(m)}_{g_1 g_2 t}=\frac{\widehat{\Delta}^{(m)}_{g_1 g_2 t}}
{\sqrt{\widehat{\mathrm{Var}}\!\left\{\widehat{\Delta}^{(m)}_{g_1 g_2 t}\right\}}},
\]
and compute the two-sided $p$-value as
$p=2\{1-\Phi(|Z^{(m)}_{g_1 g_2 t}|)\}$, where $\Phi(\cdot)$ denotes the standard normal
cumulative distribution function.

While the median difference is interpretable on the original outcome unit, it summarizes the treatment effect at a single quantile and may not fully reflect between-group differences when the outcome distributions differ in shape.
\subsection{Probability-based measure}

At a given time point $t$, we consider a probability-based measure defined as the conditional probability that a randomly selected individual from group $g_1$ has a smaller outcome value than a randomly selected individual from group $g_2$, evaluated at the average covariate profile.
Specifically, let
\begin{equation}
\Delta^{(p)}_{g_1 g_2 t}
=
\Pr\!\left(
y_{i_{g_1}t} < y_{i_{g_2}t}
\;\middle|\;
\bm{x}=\bar{\bm{x}}
\right).\label{deltap}
\end{equation}
This measure admits a direct probabilistic interpretation.
When $g_1$ corresponds to the treatment group and $g_2$ to the control group,
and smaller values of the outcome indicate better clinical status,
$\Delta^{(p)}_{g_1 g_2 t}$ can be interpreted as the probability that the treatment
produces a better outcome than the control at time point $t$,
evaluated at the overall mean covariate values.
Values of $\Delta^{(p)}_{g_1 g_2 t}$ greater than $0.5$ indicate a beneficial
treatment effect, whereas values less than $0.5$ indicate the opposite ordering.
If the direction of the outcome is reversed, the interpretation can be adjusted
accordingly by interchanging $g_1$ and $g_2$.

When the group difference is characterized solely by a shift in the location parameter, $\Delta^{(p)}_{g_1 g_2 t}$ provides an alternative summary of the treatment effect with an interpretation broadly consistent with that of the median difference.

Under the proposed BCMVR framework, the marginal distribution for each group and time point follows a power-normal distribution (PND) \cite{goto_etal_80}, which allows the probability-based measure $\Delta^{(p)}_{g_1 g_2 t}$ to be expressed as
\[
\begin{aligned}
\Delta^{(p)}_{g_1 g_2 t}
&=
\int_{0}^{\infty}
F_{\mathrm{PN}}\!\left(
y;\lambda_{g_1t},\mu_{g_1t},\sigma_{g_1t}^2
\right)
f_{\mathrm{PN}}\!\left(
y;\lambda_{g_2t},\mu_{g_2t},\sigma_{g_2t}^2
\right)
\,dy \\
&=
\int_{0}^{\infty}
y^{\lambda_{g_2t}-1}
F_{\mathrm{N}}\!\left(
z_{g_1t};\mu_{g_1t},\sigma_{g_1t}^2
\right)
f_{\mathrm{N}}\!\left(
z_{g_2t};\mu_{g_2t},\sigma_{g_2t}^2
\right)
\,dy,
\end{aligned}
\]
where $\mu_{gt}=\bm{\beta}_{gt}^{\top}\bar{\bm{x}}$ and $\sigma_{gt}^2$ denotes the
marginal variance at time point $t$, contained in $\bm{\alpha}_g$.
Here, $F_{\mathrm{PN}}(\cdot)$ and $f_{\mathrm{PN}}(\cdot)$ denote the cumulative
distribution function and probability density function of the PND, respectively, and $F_{\mathrm{N}}(\cdot)$ and $f_{\mathrm{N}}(\cdot)$
denote those of the normal distribution.

The plug-in estimator of the probability-based measure
$\Delta^{(p)}_{g_1 g_2 t}$ is obtained by evaluating the defining functional at
the maximum likelihood estimators,
\[
\widehat{\Delta}^{(p)}_{g_1 g_2 t}
=
\Delta^{(p)}_{g_1 g_2 t}
(\widehat{\bm{\theta}}_{g_1},\widehat{\bm{\theta}}_{g_2}).
\]
Let $\bm{\theta}_{g_1 g_2}
=
(\bm{\theta}_{g_1}^{\top},\bm{\theta}_{g_2}^{\top})^{\top}
$
denote the parameter vector associated with groups $g_1$ and $g_2$, and let
$\hat{V}^{(\mathrm{R})}_{\bm{\theta}_{g_1 g_2}}$ denote the corresponding
block-diagonal robust variance estimator under independence across groups.
To construct confidence intervals that respect the unit interval,
we apply the logit transformation. Let
\[
\eta^{(p)}_{g_1 g_2 t}
=
\mathrm{logit}\!\left(\Delta^{(p)}_{g_1 g_2 t}\right)
=
\log\!\left\{
\frac{\Delta^{(p)}_{g_1 g_2 t}}
{1-\Delta^{(p)}_{g_1 g_2 t}}
\right\}.
\]
The asymptotic variance of the plug-in estimator
$\widehat{\eta}^{(p)}_{g_1 g_2 t}
=
\mathrm{logit}(\widehat{\Delta}^{(p)}_{g_1 g_2 t})$
is obtained via the delta method as
\[
\widehat{\mathrm{Var}}\!\left(
\widehat{\eta}^{(p)}_{g_1 g_2 t}
\right)
=
\nabla_{\bm{\theta}_{g_1 g_2}}
\eta^{(p)}_{g_1 g_2 t}
(\widehat{\bm{\theta}}_{g_1 g_2})^{\top}
\hat{V}^{(\mathrm{R})}_{\bm{\theta}_{g_1 g_2}}
\nabla_{\bm{\theta}_{g_1 g_2}}
\eta^{(p)}_{g_1 g_2 t}
(\widehat{\bm{\theta}}_{g_1 g_2}),
\]
where the gradient is defined as
\[
\nabla_{\bm{\theta}_{g_1 g_2}}
\eta^{(p)}_{g_1 g_2 t}
(\widehat{\bm{\theta}}_{g_1 g_2})
=
\left.
\frac{\partial \eta^{(p)}_{g_1 g_2 t}(\bm{\theta}_{g_1 g_2})}
{\partial \bm{\theta}_{g_1 g_2}}
\right|_{\bm{\theta}_{g_1 g_2}=\widehat{\bm{\theta}}_{g_1 g_2}}.
\]

A Wald-type $100(1-\alpha)\%$ confidence interval for
$\Delta^{(p)}_{g_1 g_2 t}$ is obtained by back-transforming the interval for
$\eta^{(p)}_{g_1 g_2 t}$,
\[
\mathrm{logit}^{-1}\!\left(
\widehat{\eta}^{(p)}_{g_1 g_2 t}
\ \pm\
z_{1-\alpha/2}
\sqrt{
\widehat{\mathrm{Var}}\!\left(
\widehat{\eta}^{(p)}_{g_1 g_2 t}
\right)
}
\right),
\]
where $\mathrm{logit}^{-1}(u)=\{1+\exp(-u)\}^{-1}$.

To test the null hypothesis
$H_0:\Delta^{(p)}_{g_1 g_2 t}=0.5$,
we apply a Wald test on the logit scale using the statistic
\[
Z^{(p)}_{g_1 g_2 t}=\frac{
\widehat{\eta}^{(p)}_{g_1 g_2 t}
}{
\sqrt{
\widehat{\mathrm{Var}}\!\left(
\widehat{\eta}^{(p)}_{g_1 g_2 t}
\right)
}
},
\]
with the two-sided $p$-value computed from the standard normal distribution.
Details on the derivation of the gradient vectors used in the delta method are
provided in the Supporting Information.

The median difference $\Delta^{(m)}_{g_1 g_2 t}$ provides an interpretable summary
of treatment effects on the original outcome unit, but it reflects a contrast
at a single quantile.
When group differences involve distributional shape in addition to location,
the median difference alone may not fully capture differences across the entire
distribution.
In such settings, the probability-based measure $\Delta^{(p)}_{g_1 g_2 t}$ offers
a more comprehensive summary by integrating information over the full marginal
distributions.
Accordingly, while both measures are considered in this study, greater emphasis
is placed on the probability-based measure within the proposed BCMVR framework.
At the same time, differences in the central part of the distribution remain
meaningful, and the median difference can still serve as a useful supplementary
measure of treatment effects.

\subsection{Empirical small-sample adjustment}
The inference procedures for the median difference
(Section~3.1) and the probability-based measure
(Section~3.2) are derived from asymptotic theory.
In finite samples, however, Wald-type inference based on
asymptotic standard errors may exhibit slight inflation of
the type~I error rate.

To improve finite-sample performance, we applied a simple
empirical small-sample adjustment following the approach
used in Maruo \textit{et al}. \cite{maruo_etal_17}.
Specifically, the standard error was inflated by the factor
$\sqrt{n^*/(n^*-T)}$, and the resulting Wald-type statistic
was compared with a $t$ distribution with $n^*-T$
degrees of freedom, where $n^*$ denotes the total number of complete cases across the two groups, defined as participants with no missing observations at any time point.

Similar small-sample adjustments have been discussed in
the context of mixed models \cite{schluchter_elashoff_90} and are also implemented
as options for specifying the degrees of freedom in the SAS
procedure \texttt{PROC MIXED}.
In the present study, this adjustment is used as a simple
empirical correction to mitigate potential small-sample
distortions of the asymptotic Wald-type inference.
This adjustment is used as a pragmatic correction to improve finite-sample performance, 
rather than as a theoretically exact procedure.

This correction was applied to both the median difference
$\Delta^{(m)}_{g_1 g_2 t}$ and the probability-based measure
$\Delta^{(p)}_{g_1 g_2 t}$.
The impact of this adjustment on type~I error control and
statistical power is examined in the simulation study in
Section~4.
\section{Simulation and Empirical Evaluation}

\begin{figure*}[b!]
\centerline{\includegraphics{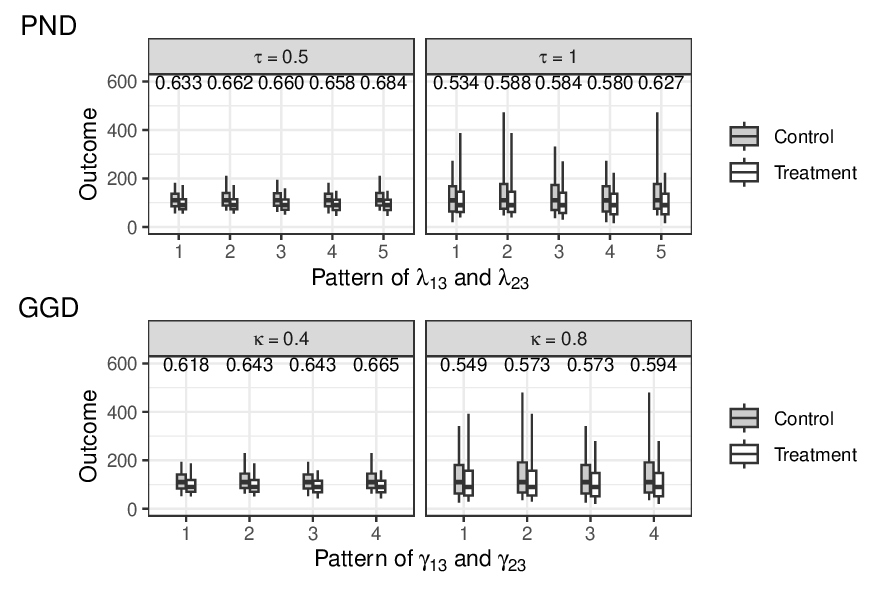}}
  \caption{Conditional distributions at the final time point ($t=T$) under the alternative hypothesis. Values above are probability indices.}
  \label{fig:boxplot}
\end{figure*}
\subsection{Simulation design}

We considered a randomized, parallel-group comparative trial with two treatment groups and repeated measurements of a continuous outcome, where higher values indicate worse disease status.
Participants were randomized to either a control group ($g=1$) or a treatment group ($g=2$), and the outcome was measured at $T$ time points indexed by $t=0, 1,\ldots,T$, where $t=0$ meant baseline visit and $T=3$.
Sample sizes were balanced between groups, with $N_1=N_2\in\{25,\,50,\,100\}$.

\noindent\textbf{Correctly specified marginal distributions (PND).}
As a baseline scenario, the marginal distribution at each time point was assumed to follow a PND under the reparameterization of Maruo \textit{et al}. \cite{maruo_etal_17_cs}.
Specifically, for participant $i$ in group $g\in\{1,2\}$ at time $t=0, 1,\ldots,T$, the outcome was generated as
\[
y_{igt} \sim \mathrm{PND}(\lambda_{gt}, \xi_{gt}, \tau),
\]
where $\lambda_{gt}$ is the power (shape) parameter, $\xi_{gt}$ is the median, and $\tau$ is a scale parameter defined as the ratio of the interquartile range to the median, with $\tau\in\{0.5,\,1\}$.

The power parameter was specified to evolve linearly over time according to $\lambda_{gt} = (t/T)\lambda_{gT}$, with $\lambda_{g0}=0$ at baseline.
Under the null hypothesis, the terminal values were identical across groups,
\[
\lambda_{1T} = \lambda_{2T} \in \{-0.5,\,0,\,0.5\},
\]
whereas under the alternative hypothesis, the following five configurations were considered:
\[
(\lambda_{1T},\lambda_{2T}) \in
\{(0.5,-0.5),\,(-0.5,-0.5),\,(0,0),\,(0.5,0.5),\,(-0.5,0.5)\}.
\]
For ease of reference, these configurations were labeled as Patterns~1--5, respectively.

The median trajectory in the PND setting was specified as $\xi_{gt} = 100 \pm 10(t/T)$.
Under the null hypothesis, both groups shared the same positive median trajectory, whereas under the alternative hypothesis the control group ($g=1$) followed the positive trajectory and the treatment group ($g=2$) followed the negative trajectory.

\noindent\textbf{Misspecified marginal distributions (GGD).}
To assess robustness to model misspecification, we additionally considered scenarios in which the true marginal distribution followed a generalized gamma distribution (GGD)\cite{prentice_74}.
Outcomes were generated as
\[
y_{igt} \sim \mathrm{GGD}(\gamma_{gt}, \nu_{gt}, \kappa),
\]
where $\gamma_{gt}$ is a shape parameter, $\nu_{gt}$ is a location parameter, and $\kappa$ is a scale parameter, with $\kappa\in\{0.4,\,0.8\}$.

The shape parameter evolved over time as $\gamma_{gt} = (t/T)\gamma_{gT}$.
Under the null hypothesis,
\[
\gamma_{1T} = \gamma_{2T} \in \{-0.5,\,0.5\},
\]
whereas under the alternative hypothesis,
\[
(\gamma_{1T},\gamma_{2T}) \in
\{(0.5,-0.5),\,(-0.5,-0.5),\,(0.5,0.5),\,(-0.5,0.5)\}.
\]
These configurations were labeled as Patterns~1--4.

When $\gamma_{gt}=0$, the GGD reduces to a log-normal distribution, which is also contained within the PND family; therefore, this case was not included in the GGD scenarios. Only the baseline followed a log-normal distribution.

In the GGD setting, the median trajectory was specified as $\xi_{gt} = 100 \pm 15(t/T)$, and the group-specific median trajectories under the null and alternative hypotheses were specified in the same manner as in the PND setting.
For each $(g,t)$, the location parameter $\nu_{gt}$ was determined so that the median of the GGD equaled $\xi_{gt}$.

\noindent\textbf{Illustration of conditional distributions and treatment effects.}
Figure~\ref{fig:boxplot} displays the distributions at the final time point ($t=T$) under the alternative hypothesis for each shape-parameter pattern and scale parameter setting, evaluated at a representative baseline value.

The shape-parameter patterns were designed to represent qualitatively different forms of treatment effects beyond simple location shifts.
In Pattern~1, the treatment group exhibits a more favorable median outcome; however, its distribution has a heavier upper tail, indicating that the treatment effect diminishes or may even become unfavorable among more severe cases.
In contrast, in the final pattern (Pattern~5 for PND and Pattern~4 for GGD), the treatment effect is reflected not only in an improved median but also in a lighter upper tail, indicating an amplified benefit among more severe cases.
In the remaining patterns, the shape parameters at the final time point are identical across groups, so that between-group differences are primarily driven by location rather than distributional shape.

Except for Pattern~3 in the PND scenarios, the shape parameters vary over time, and thus the assumption of shape invariance across time points underlying BCMMRM is violated.
Pattern~3 corresponds to the special case $\lambda_{gt}=0$, for which the distribution reduces to a log-normal distribution at all time points.

\noindent\textbf{Correlation structure.}
Within-participant longitudinal dependence was characterized by an AR(1) structure with correlation parameter $\rho=0.7$.
In the PND setting, this correlation structure is inherent to the definition of the multivariate PND on the transformed scale.
In the GGD setting, dependence was introduced via a multivariate normal copula with AR(1) correlation, which was subsequently transformed to obtain GGD marginals.

\noindent\textbf{Missing-data mechanism.}
Dropout due to insufficient treatment effect was incorporated through a logistic regression--based missing-at-random (MAR) mechanism.
The dropout probability depended on the most recent observed outcome: on the Box--Cox transformed and scaled (unit standard deviation) outcome in the PND setting, and on the scaled outcome without transformation in the GGD setting.
The intercept parameter was calibrated so that the overall dropout rate across both groups was approximately 30\% at the final time point.
Because the dropout mechanism depended only on observed past outcomes, the MAR assumption was satisfied.
Missing values were generated solely through dropout, resulting in a monotone missing data pattern.
In addition, for reference, we have also set a scenario with 0\% dropout.

\noindent\textbf{Analysis procedures.}
Each simulation scenario was replicated $10{,}000$ times.
For each simulated dataset, a conventional MMRM assuming normality (used for hypothesis testing only), BCMMRM, BCMVR based on the median difference [BCMVR(M)], and BCMVR based on the probability-based measure [BCMVR(P)] were applied.
In all analyses, baseline was included as a covariate. For the conventional MMRM, the baseline outcome on the original scale was included, together with its interaction with time. An unstructured covariance matrix was assumed for the repeated measurements, and the Kenward--Roger method \cite{kenward_roger_97} was used for degree-of-freedom adjustment. For BCMMRM and BCMVR, the baseline outcome after the Box--Cox transformation was included as a covariate, and for BCMMRM its interaction with time was also included. This specification was adopted to maintain consistency with the scale on which each model was formulated. In particular, BCMMRM and BCMVR model the transformed outcome, and therefore including the transformed baseline provides a covariate representation aligned with the transformed-scale model. In the simulation settings, the baseline covariate followed a log-normal distribution, so that the inverse transformation of its mean on the transformed scale coincided with its median on the original scale. Thus, conditioning on the mean of the transformed baseline corresponded to conditioning on the median of the baseline on the original scale.
In addition, their small-sample adjusted versions described in Section~3.3 were also applied, denoted as BCMVR(M)a and BCMVR(P)a.

For BCMMRM, BCMVR(M), and BCMVR(M)a, inference was performed for the median difference at the final time point, including a two-sided hypothesis test of
\[
H_0:\Delta^{(m)}_{g_1 g_2 T}=0
\]
at significance level $\alpha=0.05$ and construction of a $95\%$ confidence interval.

For BCMVR(P) and BCMVR(P)a, inference was performed for the probability-based treatment effect at the final time point, including a two-sided hypothesis test of
\[
H_0:\Delta^{(p)}_{g_1 g_2 T}=0.5
\]
at significance level $\alpha=0.05$ and construction of a $95\%$ confidence interval.

\noindent\textbf{Performance measures.}
Simulation performance was evaluated in terms of empirical type~I error rates under the null hypothesis and empirical power under the alternative hypothesis, empirical bias of the estimated treatment effect, and empirical coverage probabilities of the nominal $95\%$ confidence intervals.
Furthermore, we evaluated the proportion of simulations in which the BCMVR model was selected over the BCMMRM model based on a likelihood ratio test, where BCMMRM and BCMVR correspond to the null and alternative hypotheses, respectively, at a significance level of 0.05. This comparison serves as a diagnostic for detecting differences in distributional shape between treatment groups.
As an additional measure of estimation accuracy, we evaluated the ratio of the average estimated standard errors to the standard deviation of estimated treatment effects across simulation replicates, multiplied by 100. Values close to 100 indicate small bias in the standard errors, whereas values below and above 100 indicate underestimation and overestimation of the standard errors, respectively.
In addition, we recorded the convergence rate of the estimation algorithm for the BCMVR model in each simulation scenario.
For all performance measures other than the convergence rate, simulation replicates in which the BCMVR estimation algorithm failed to converge were excluded from the evaluation.

\noindent\textbf{Software.}
All simulations were conducted using R (version 4.6.0; R Core Team, Vienna, Austria) and several R packages. The MMRM analyses were performed using the \texttt{mmrm} package \cite{mmrm_pkg}, while the BCMMRM analyses were conducted using the \texttt{bcmixed} package \cite{bcmixed}. The proposed BCMVR method was implemented using the \texttt{bcmvr} package \cite{bcmvr}. Multivariate power-normal random variables were generated using the \texttt{powerNormal} package \cite{pnd}, and generalized gamma random variables were generated using the \texttt{flexsurv} package \cite{flexsurv}.

\subsection{Simulation results}
In this section, we present the simulation results under the 30\% dropout setting. Results for the no-dropout setting, together with additional simulation results (SE ratio, coverage), are provided in the Supporting Information.

\begin{table}[t!]
\caption{Proportion of simulations in which the BCMVR model was selected over the BCMMRM model based on the likelihood ratio test.
``Null scenarios'' and ``Alternative scenarios'' refer to simulation settings with and without treatment effects, respectively.
Under the null scenarios, the columns $-0.5$, $0$, and $0.5$ represent the common shape parameter values at time 3.
}
\label{tab:lrt}
\centering
\small
\setlength{\tabcolsep}{4pt}

\begin{tabular}{
lcc
S[table-format=2.1]
S[table-format=2.1]
S[table-format=2.1]
S[table-format=2.1]
S[table-format=2.1]
S[table-format=2.1]
S[table-format=2.1]
S[table-format=2.1]
}
\hline

Dist. & Scale & $N_g$ &
\multicolumn{3}{c}{Null scenarios} &
\multicolumn{5}{c}{Alternative scenarios} \\

\cline{4-6} \cline{7-11}

& & & {$-0.5$} & {$0$} & {$0.5$} &
{1} & {2} & {3} & {4} & {5} \\

\hline

PND & 0.5 & 25  &  9.6 &  9.0 & 10.6 & 19.3 & 10.2 &  8.9 & 10.7 & 17.5 \\
 &  & 50  & 10.7 &  7.7 & 11.2 & 33.4 & 11.1 &  7.3 & 11.8 & 31.5 \\
&  & 100 & 13.2 &  6.1 & 15.0 & 65.0 & 15.1 &  5.8 & 17.1 & 62.8 \\

 & 1.0 & 25  & 19.2 & 12.3 & 19.3 & 61.5 & 19.3 & 12.7 & 19.9 & 60.1 \\
 &  & 50  & 22.3 &  7.7 & 22.9 & 91.4 & 23.1 &  7.2 & 24.1 & 90.7 \\
 &  & 100 & 40.8 &  6.1 & 44.6 & 99.9 & 43.5 &  6.2 & 47.6 & 99.9 \\

\hline

GGD & 0.4 & 25  & 12.0 & \multicolumn{1}{c}{--} & 11.8 &
18.3 & 11.7 & 12.8 & 17.8 & \multicolumn{1}{c}{--} \\

 &  & 50  & 10.1 & \multicolumn{1}{c}{--} &  9.5 &
26.3 & 10.8 & 10.5 & 25.4 & \multicolumn{1}{c}{--} \\

 &  & 100 & 11.2 & \multicolumn{1}{c}{--} & 11.7 &
50.6 & 12.0 & 12.2 & 50.4 & \multicolumn{1}{c}{--} \\

 & 0.8 & 25  & 15.2 & \multicolumn{1}{c}{--} & 14.7 &
21.4 & 15.1 & 14.6 & 21.0 & \multicolumn{1}{c}{--} \\

 &  & 50  & 10.0 & \multicolumn{1}{c}{--} &  9.9 &
25.8 & 10.7 & 10.9 & 25.7 & \multicolumn{1}{c}{--} \\

 &  & 100 & 11.2 & \multicolumn{1}{c}{--} & 11.0 &
49.6 & 12.2 & 11.2 & 50.1 & \multicolumn{1}{c}{--} \\
\hline
\end{tabular}
\end{table}

\noindent\textbf{Model selection between BCMMRM and BCMVR.}
Table~\ref{tab:lrt} shows the proportion of simulations in which the
BCMVR model was selected over the BCMMRM model based on the likelihood
ratio test comparing the BCMMRM model (null hypothesis) and the BCMVR
model (alternative hypothesis). We first describe the results under the
alternative scenarios with treatment effects.

Under the PND data-generating mechanisms, the BCMVR model was selected
more frequently in patterns 1 and 5, where the shape parameter
$\lambda_{gt}$ differed between treatment groups as well as across time
points. In contrast, the selection proportion was low in patterns
2--4, where $\lambda_{gt}$ differed only across time points but not between
groups. In addition, the selection proportion increased as the scale
parameter $\tau$ became larger.

Under the GGD scenarios, a similar dependence on the shape parameter
patterns was observed, with higher selection proportions in patterns
1 and 4 where the shape parameter $\gamma_{gt}$ differed between treatment
groups. In contrast to the PND scenarios, the scale parameter
$\kappa$ had little impact on the model selection results.

Under the null scenarios with no treatment effect, the selection
proportions were broadly similar to those observed in scenarios 2--4
for the PND settings and scenarios 2--3 for the GGD settings.

These results suggest that the likelihood ratio test can detect
model misspecification caused by group-specific differences in
distributional shape, although the detection ability becomes
limited when the sample size is small.
In scenarios where the
BCMMRM model was correctly specified, the empirical type~I error
rates approached the nominal significance level as the sample size
increased.

In scenarios without missing data, the selection rate of BCMVR increased under the alternative hypothesis, while under the null hypothesis it became closer to the nominal significance level (Table S1).

\begin{figure*}[b!]
\centerline{\includegraphics{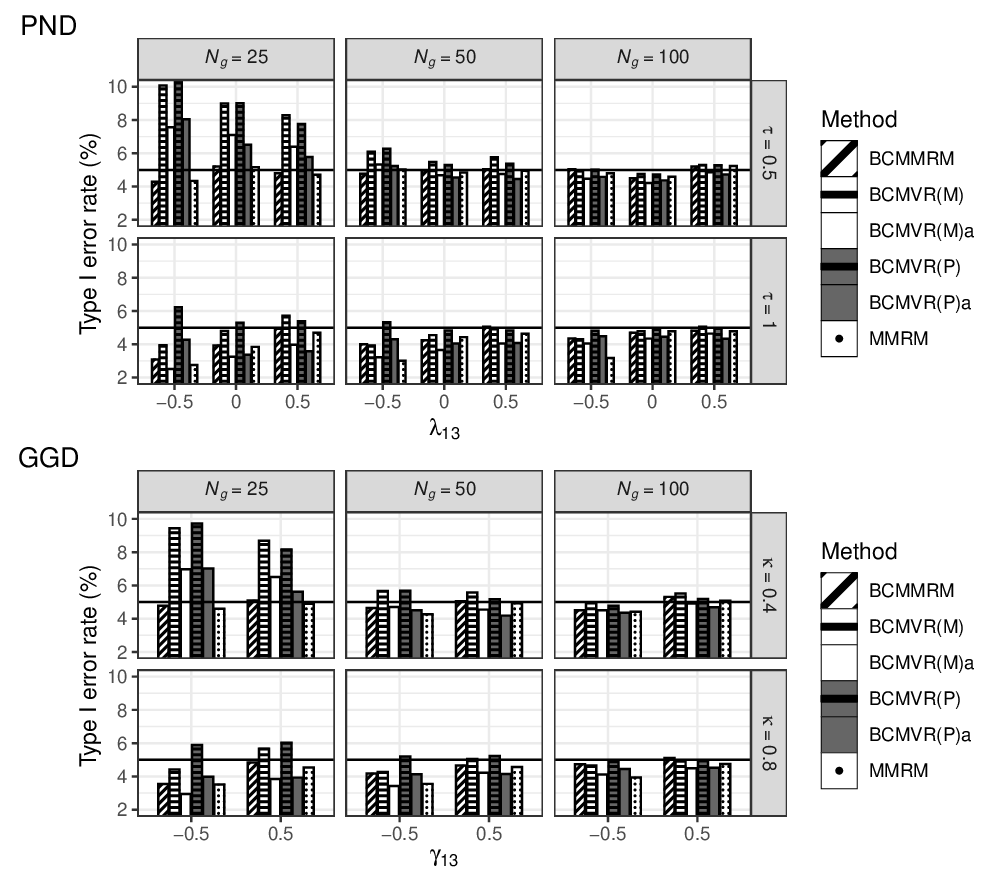}}
  \caption{Empirical type I error of tests for treatment effect..}
  \label{fig:test_size}
\end{figure*}

\noindent\textbf{Empirical type I error.}
The empirical type~I error rates are shown in
Figure~\ref{fig:test_size}.
Inflation of the type~I error rate was observed when the scale
parameters were small (i.e., small $\tau$ in PND and small $\kappa$
in GGD) and the sample size per group was $N_g=25$ for the proposed methods.
In these settings, the degree of inflation was substantial and the
empirical small-sample adjustment described in Section~3.3 was not
sufficient to fully control the nominal level.
In the other scenarios, however, the small-sample adjustment
maintained the nominal significance level. In contrast, the asymptotic method showed slight inflation when the sample size
per group was $N_g=50$.
Based on these results, the following comparisons focus on the
small-sample adjusted procedures.
\begin{table}[t]
\centering
\caption{Empirical convergence probabilities (\%) of the estimation algorithm across simulation scenarios.}
\label{tab:conv}
\begin{tabular}{llrrrrr}
\toprule
Dist. & Scenario & Min. & Q1 & Median & Q3 & Max. \\
\midrule
PND & $N_g=25$, $\tau=0.5$ & 87.3 & 89.9 & 92.2 & 93.8 & 96.3 \\
PND & Other                & 99.2 & 99.8 & 100.0 & 100.0 & 100.0 \\
GGD & $N_g=25$, $\kappa=0.4$ & 91.3 & 91.8 & 93.6 & 95.2 & 95.7 \\
GGD & Other                & 99.4 & 99.8 & 100.0 & 100.0 & 100.0 \\
\bottomrule
\end{tabular}
\end{table}
Summary statistics of the convergence rates of the estimation
algorithm for the BCMVR model across all simulation scenarios, including those generated
under the alternative hypothesis, are presented in
Table~\ref{tab:conv}.
Lower convergence rates were observed when $N_g=25$ and the scale
parameters were small (i.e., $\tau=0.5$ for PND and $\kappa=0.4$ for
GGD).
In the other scenarios, the algorithm converged in almost all
simulation runs.
Notably, these settings largely coincide with those in which
inflation of the type~I error rate was observed.

In scenarios without missing data, the inflation of the type I error rate for BCMVR observed in small-sample settings was attenuated (Figure S6).

\begin{figure*}[b!]
\centerline{  \includegraphics{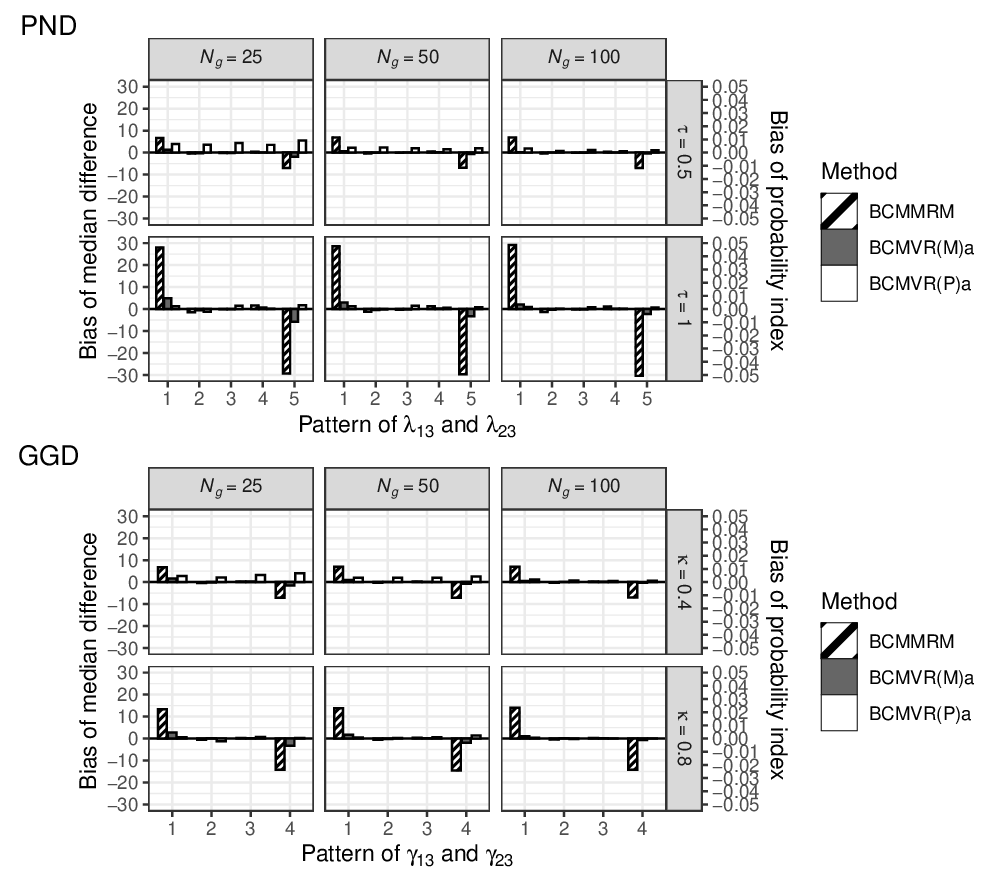}}
\caption{Bias of treatment effect under the alternative hypothesis.
Biases of the median difference for BCMMRM and BCMVR(M)a are shown on the left vertical axis, while biases of the probability-based measure for BCMVR(P)a are shown on the right vertical axis.}
  \label{fig:bias_a}
\end{figure*}

\noindent\textbf{Empirical bias of treatment effect.}
The bias of treatment effect under the alternative hypothesis is shown in Figure~\ref{fig:bias_a}.
For the BCMMRM method, little bias was observed in scenarios where
the shape parameters differed only across time points but not between
treatment groups (PND patterns 2 and 4, and GGD patterns 2 and 3).
In contrast, when the shape parameters differed between treatment
groups at the time point of interest (PND patterns 1 and 5, and GGD
patterns 1 and 4), noticeable bias was observed.

In comparison, the BCMVR method showed negligible bias for both the
median difference and the probability-based measure across all
scenarios, including those generated from the GGD model where the
BCMVR model was misspecified.

\begin{figure*}[b!]
\centerline{  \includegraphics{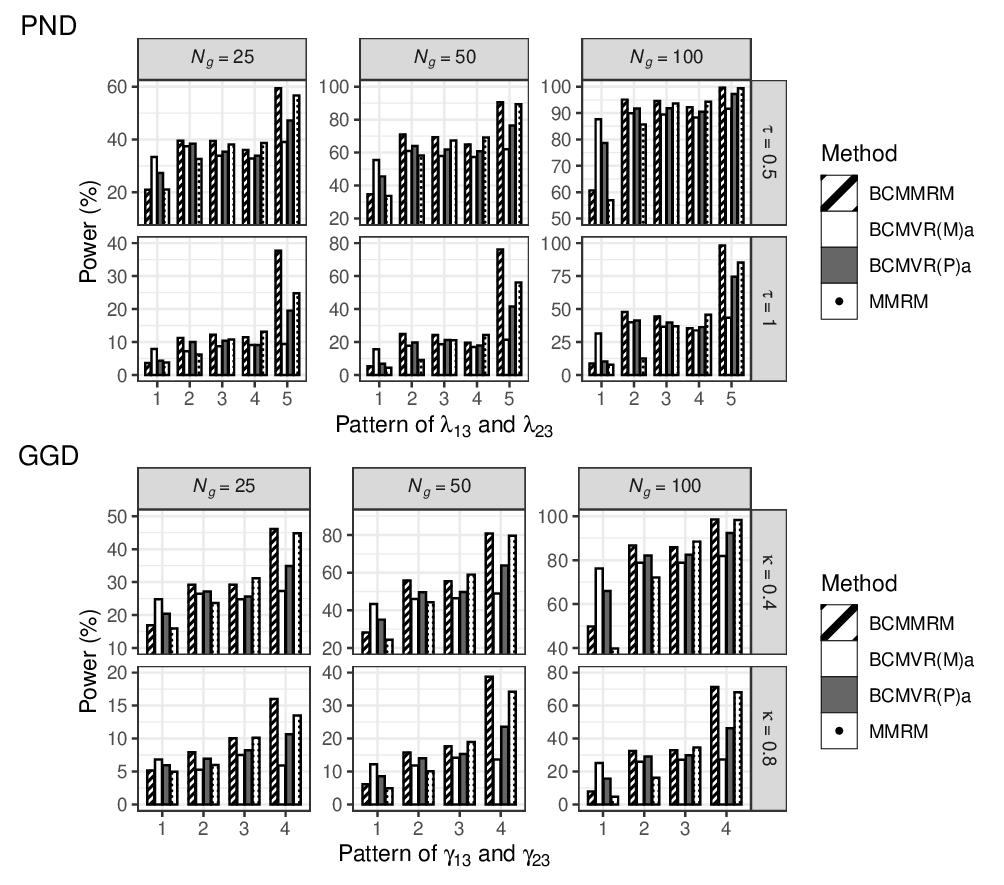}}
\caption{Empirical power of tests for treatment effect.}
  \label{fig:power}
\end{figure*}

Under the null hypothesis, virtually no bias was observed in any situation or method (Figure S1).

In scenarios without missing data, convergence rates were slightly improved (Table S2).

\noindent\textbf{Empirical power.}
The empirical power of the tests is shown in
Figure~\ref{fig:power}.
The BCMMRM method generally showed high power across many scenarios.
However, this tendency partly reflects the bias observed in the
estimated treatment effects when the shape parameters differed
between treatment groups.

The conventional MMRM achieved relatively high power in scenarios with small sample sizes ($N_g = 25$) and/or mild skewness (e.g., $\lambda_{gT} = 0.5$ or $\gamma_{gT} = 0.5$). However, in other scenarios, the power was rather reduced.

The BCMVR(M) method showed relatively stable power across scenarios,
although the power was generally lower than that of the BCMMRM
method.

The BCMVR(P) method showed power levels between those of BCMMRM and
BCMVR(M), indicating a moderate trade-off between robustness and
efficiency.

Although statistical power generally increased in scenarios without missing data, the relative differences between the methods were similar to those observed in the presence of missing data (Figure S7).

\noindent\textbf{SE ratio.}
The ratios of the model-based standard errors to the empirical standard deviations are presented in Figures S2--S3. For both BCMVR methods, standard errors tended to be underestimated when the sample size was small ($N_g=25$) and the scale parameter was small. This finding was consistent with the observed inflation of the type I error rate under these settings. 

For the generalized gamma distribution with $N_g=25$ and $\kappa=0.8$, some underestimation was observed for the standard errors of BCMVR(M). However, this was primarily attributable to the heavy-tailed distribution of the median difference estimator, which inflated its empirical standard deviation, and therefore had little impact on the observed type I error rate. Apart from these settings, the standard error estimates were generally unbiased. 

In contrast, BCMMRM exhibited a slight tendency to underestimate standard errors in scenarios where the distributional shapes differed between treatment groups.

When no missing data were present, the bias of SE for the proposed methods was attenuated (Figures S10--S11).

\noindent\textbf{Empirical coverage probability.}
The empirical coverage probabilities of the nominal $95\%$ confidence
intervals are shown in Supplementary Figures~S4-5. For the
BCMMRM method, the coverage probabilities tended to fall below the
nominal level in scenarios where noticeable bias was observed in the
estimated treatment effects, as expected from the bias results.

In contrast, the proposed BCMVR methods generally maintained coverage
close to the nominal level across most scenarios. Under the smallest
sample size ($N_g=25$) with small scale parameters, the coverage
probabilities were slightly below the nominal level, which is
consistent with the type~I error inflation observed in these settings.

In the absence of missing data, the empirical coverage probabilities of the proposed methods were closer to the nominal level (Figures S12--S13).

\subsection{Application to ACTG 193A trial data}
To illustrate the proposed methods, we analyzed data from the AIDS Clinical Trial Group (ACTG) 193A study \cite{henry_etal_98, fitzmaurice_etal_11} which is available in the \texttt{bcmixed} R package. The study was a randomized clinical trial evaluating antiretroviral treatment strategies in patients with advanced HIV infection. CD4 cell counts were measured at weeks 8, 16, 24, and 32.

In the present analysis, we considered two treatment groups: the two-drug regimen (zidovudine + didanosine), denoted by $g=1$, and the three-drug regimen (zidovudine + didanosine + nevirapine), denoted by $g=2$. The comparison focuses on evaluating the add-on effect of nevirapine.

The sample sizes were $N_1=293$ and $N_2=308$ for the two-drug and three-drug groups, respectively. At week 32, the numbers of observed outcomes were 187 (63.8\%) in the two-drug group and 208 (67.5\%) in the three-drug group.

\begin{figure*}[b!]
\centerline{\includegraphics[width=0.72\textwidth]{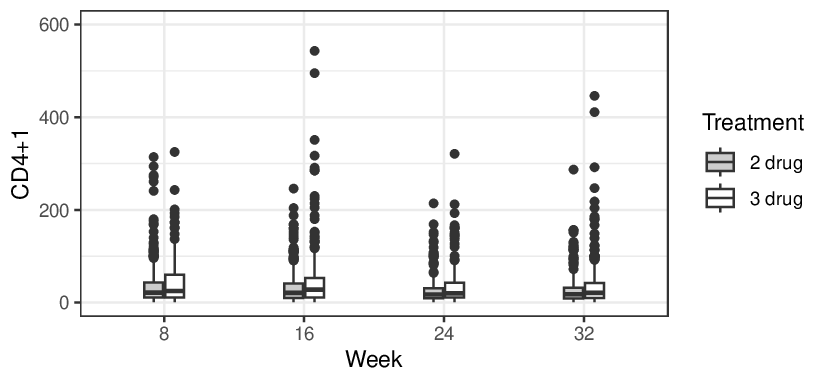}}
\caption{Distributions of CD4 cell counts (+1 shifted) at weeks 8, 16, 24, and 32 in the two-drug ($g=1$) and three-drug ($g=2$) groups in the ACTG 193A study.}
  \label{fig:actg_box}
\end{figure*}

Figure~\ref{fig:actg_box} shows the distributions of CD4 cell counts (shifted by adding 1 to avoid zero values) at weeks 8, 16, 24, and 32 in the two treatment groups. The distributions appear markedly right-skewed at all visits, supporting the use of transformation-based modeling approaches and distribution-sensitive treatment effect summaries.

We applied BCMMRM, BCMVR(M), and BCMVR(P) to estimate the treatment effect at week 32. As in the simulation study, the baseline outcome after the Box--Cox transformation was included as a covariate.
In the simulation studies and throughout this paper, smaller outcome values are assumed to represent better clinical status. However, for the CD4 cell count data, larger values indicate better clinical status. Therefore, the direction of inequality in equation (\ref{deltap}) is reversed in this application.

The likelihood ratio test comparing BCMMRM and BCMVR yielded a $p$-value of 0.021, suggesting potential differences in distributional shape between the treatment groups. The estimated shape parameter under BCMMRM was $\lambda=0.134$. Under BCMVR, the estimated shape parameters were $\bm{\lambda}_1=(0.119,\,0.208,\,0.192,\,0.190)^{\top}$ for the two-drug group ($g=1$) and $\bm{\lambda}_2=(0.143,\,0.082,\,0.163,\,0.076)^{\top}$ for the three-drug group ($g=2$). These estimates suggest some differences in distributional shape between the two treatment groups.

BCMMRM estimated a median difference of 6.46 (95\% CI: 3.16 to 9.76, $p<0.001$). BCMVR(M) estimated a median difference of 4.89 (95\% CI: 1.54 to 8.25, $p=0.004$). The probability-based measure BCMVR(P) estimated that the probability that a randomly selected patient in the three-drug group has a higher CD4 count than a patient in the two-drug group was 0.585 (95\% CI: 0.534 to 0.635, $p=0.001$).

Overall, the BCMVR framework provides complementary perspectives on treatment effects through distributional summaries while accommodating potential differences in distributional shape between treatment groups.

\section{Discussion}
In this study, we proposed the BCMVR framework for treatment effect inference in longitudinal clinical trial data with potentially skewed outcome distributions. The proposed approach extends the conventional BCMMRM framework by enabling inference based on interpretable distributional summaries, including the median difference and a probability-based treatment effect measure.

The simulation results demonstrated that the conventional BCMMRM approach can produce biased estimates of treatment effects when the underlying outcome distributions differ in shape between treatment groups. In contrast, the BCMVR procedures provided nearly unbiased estimation across all considered scenarios, including those under model misspecification with generalized gamma data-generating mechanisms. These findings suggest that BCMVR offers a more robust alternative when distributional heterogeneity exists.

Regarding hypothesis testing, BCMMRM generally exhibited high statistical power; however, this was partly attributable to bias under model misspecification. The BCMVR(M) procedure showed relatively stable but moderate power, whereas BCMVR(P) tended to achieve power levels between those of BCMMRM and BCMVR(M), suggesting a useful balance between robustness and efficiency.

The likelihood ratio test comparing BCMMRM and BCMVR provided informative diagnostics for detecting distributional misspecification. In scenarios with differing shape parameters between treatment groups, the BCMVR model was selected more frequently. However, the selection proportion was modest in small-sample settings, likely reflecting estimation instability and reduced convergence associated with the increased model complexity of BCMVR. Consistent with this, the simulation results showed reduced estimation accuracy and lower convergence rates when both the sample size and scale parameters were small. In practice, the advantages of BCMVR may become more evident when the sample size per group is approximately 50 or larger.

Taken together, these results suggest that BCMVR(P) can be recommended as a practical default procedure. The probability-based treatment effect provides a favorable balance between robustness and efficiency while remaining clinically interpretable. However, when the sample size is small and substantial differences in distributional shape are unlikely, the conventional BCMMRM approach may still offer more stable inference due to its simpler structure.

Importantly, the BCMVR framework should not be viewed merely as an exploratory tool but as a unified inferential framework based on distributional summaries of longitudinal outcomes. In particular, the probability-based measure integrates information over the entire marginal distributions and can capture differences beyond central tendency. In skewed distributions, the mean may provide a limited representation of the outcome, and even the median may fail to adequately reflect group differences when distributional shapes differ between treatment groups. In such settings, the probability-based measure offers a more comprehensive assessment of treatment effects. The proposed framework is related to probabilistic indices such as the Mann--Whitney parameter and the probability of superiority, but extends these ideas to longitudinal clinical trial settings with potentially skewed outcomes and missing observations under MAR. By combining the Box--Cox transformation with a parametric multivariate regression framework, BCMVR provides a flexible yet unified modeling approach that enables covariate-adjusted inference for both median-based and probability-based treatment effects within the same longitudinal framework.

From a clinical perspective, such distributional summaries may offer intuitive measures of treatment benefit. Depending on the direction of the outcome, the probability-based measure can be interpreted as the probability that a randomly selected patient from the treatment group has a better outcome than a randomly selected patient from the control group. This interpretation may be easier to communicate to clinicians than summaries based solely on a single summary measure such as the mean or median.

To facilitate practical implementation, we provide an R package, \texttt{bcmvr}, implementing the proposed framework. The package is available at \url{https://github.com/kzkzmr/bcmvr}. Detailed usage and examples are provided in the GitHub repository and in the package documentation (accessible via \texttt{?bcmvr} in R). The package can be installed using standard tools (e.g., \texttt{remotes::install\_github("kzkzmr/bcmvr")}).

The choice of the Box--Cox transformation and the PND can be justified from both empirical and theoretical perspectives. Previous work \cite{maruo_etal_17_cs} has shown that the PND provides a good fit across a wide range of clinical laboratory measurements.
Furthermore, the Box--Cox formulation ensures that the conditional distribution of the outcome given covariates remains within a tractable family (PND), enabling closed-form or easily computable expressions for clinically interpretable quantities such as median differences and probability-based measures. This property stems from the fact that, after transformation, the outcome follows a multivariate normal model, whose conditional distribution is well characterized through linear structures. In this sense, the approach effectively leverages the tractability of the normal distribution under linear operations. In contrast, models that specify a particular marginal distribution---such as the GGD---do not generally preserve this closure property, as the corresponding conditional distribution given covariates typically falls outside the assumed family, thereby complicating inference for such measures. This structural advantage is particularly important when inference targets are defined through conditional distributions, rather than marginal summaries alone.
Importantly, the proposed framework is not restricted to the PND but can be extended to other flexible distributional families, highlighting that its primary contribution lies in the inferential framework rather than in the specific choice of distribution.

Several limitations should be noted. First, the simulation settings considered only a limited range of longitudinal structures and outcome distributions. Broader evaluations under more complex designs and additional distributional families would further clarify the operating characteristics of the proposed methods. In addition, because BCMVR introduces group- and time-specific transformation parameters, estimation may become unstable in small samples. In particular, performance deteriorated when both the sample size and scale parameters were small, as reflected in type~I error inflation and reduced convergence rates. One possible explanation is that the flexibility of BCMVR increases the number of nuisance parameters relative to the available information. In the present setting, the transformed outcomes may exhibit time-specific variances induced by the group- and time-specific transformation parameters, making simple covariance structures such as compound symmetry difficult to justify. Consequently, more flexible covariance structures, such as unstructured covariance matrices, are often required, potentially leading to unstable covariance estimation in small samples. Furthermore, the first-order approximation underlying the delta method may be insufficient for nonlinear treatment effect measures, particularly when dropout further reduces the effective information size.
Another limitation is that the proposed treatment effect is defined conditionally on covariates being fixed at their overall sample means. In randomized clinical trials, marginal or covariate-standardized probability-based estimands may also be of interest because they provide population-level summaries of treatment effects. Extending the proposed framework to accommodate such marginal interpretations represents an important direction for future research. Furthermore, although the present study focused on randomized clinical trials, the proposed framework may also be applicable to observational studies through appropriate adjustment for confounding covariates. Investigation of causal and marginal interpretations of the probability-based measure in observational settings is another important topic for future work.

Overall, the BCMVR framework extends conventional MMRM and BCMMRM methodologies by enabling treatment effect inference based on interpretable distributional summaries while accommodating differences in distributional shape. This provides a flexible approach for analyzing longitudinal clinical trial data when outcome distributions deviate from standard normality assumptions.

\section*{Funding}

This work was supported by JSPS KAKENHI Grant Numbers 23K11003 and 26K02873.

\section*{Author Contributions}

Kazushi Maruo conceived the study, developed the methodology, conducted the analyses, and drafted the manuscript. 
Ryota Ishii and Yusuke Yamaguchi reviewed the manuscript and verified the analytical results. 
Toshio Shimokawa, Tomoyuki Sugimoto, and Masahiko Gosho reviewed the manuscript and provided supervision. 
All authors approved the final version of the manuscript.


\section*{Financial Disclosure}

None reported.

\section*{Conflicts of Interest}

The authors declare no conflicts of interest.

\bibliographystyle{unsrt}
\bibliography{bibfile}

@article{box_cox_64,
  author  = {G. E. P. Box and D. R. Cox},
  title   = {An Analysis of Transformations},
  journal = {Journal of the Royal Statistical Society, Series B},
  year    = {1964},
  volume  = {26},
  number  = {2},
  pages   = {211--252},
  doi     = {10.1111/j.2517-6161.1964.tb00553.x}
}

@book{fitzmaurice_etal_11,
  author    = {G. M. Fitzmaurice and N. M. Laird and J. H. Ware},
  title     = {Applied Longitudinal Analysis},
  edition   = {2},
  year      = {2011},
  publisher = {Wiley},
  address   = {New York},
  doi       = {10.1002/9781119513469}
}

@article{henry_etal_98,
  author  = {K. Henry and A. Erice and C. Tierney and H. H. {Balfour Jr.} and M. A. Fischl and A. Kmack and S. H. Liou and A. Kenton and M. S. Hirsch and J. Phair and A. Martinez and J. O. Kahn},
  title   = {A randomized, controlled, double-blind study comparing the survival benefit of four different reverse transcriptase inhibitor therapies (three-drug, two-drug, and alternating drug) for the treatment of advanced {AIDS}. {AIDS Clinical Trial Group 193A Study Team}},
  journal = {Journal of Acquired Immune Deficiency Syndromes and Human Retrovirology},
  year    = "1998",
  volume  = "19",
  number  = "4",
  pages   = "339--349",
  doi     = {10.1097/00042560-199812010-00004}
}

@article{mallinckrodt_etal_01,
  author  = {C. H. Mallinckrodt and W. S. Clark and S. R. David},
  title   = {Accounting for dropout bias using mixed-effects models},
  journal = {Journal of Biopharmaceutical Statistics},
  year    = {2001},
  volume  = {11},
  number  = {1--2},
  pages   = {9--21},
  doi     = {10.1081/BIP-100104194}
}

@article{maruo_etal_17,
  author  = {K. Maruo and Y. Yamaguchi and H. Noma and M. Gosho},
  title   = {Interpretable inference on the mixed effect model with the {Box}--{Cox} transformation},
  journal = {Statistics in Medicine},
  year    = {2017},
  volume  = {36},
  number  = {15},
  pages   = {2420--2434},
  doi     = {10.1002/sim.7279}
}

@article{maruo_etal_17_cs,
  author  = {K. Maruo and T. Yamabe and Y. Yamaguchi},
  title   = {Statistical simulation based on right skewed distributions},
  journal = {Computational Statistics},
  year    = {2017},
  volume  = {32},
  pages   = {889--907},
  doi     = {10.1007/s00180-016-0664-4}
}

@article{goto_etal_80,
  author  = {M. Goto and T. Inoue},
  title   = {Some properties of the power normal distribution},
  journal = {Japanese Journal of Biometrics},
  year    = {1980},
  volume  = {1},
  pages   = {28--54},
  doi     = {10.5691/jjb.1.28}
}

@article{bcmixed,
  author  = {K. Maruo and R. Ishii and Y. Yamaguchi and M. Gosho},
  title   = {bcmixed: {A} Package for Median Inference on Longitudinal Data with the {Box}--{Cox} Transformation},
  journal = {The R Journal},
  year    = {2021},
  volume  = {13},
  number  = {2},
  pages   = {253--265},
  doi     = {10.32614/RJ-2021-083},
  issn    = {2073-4859}
}

@article{schluchter_elashoff_90,
  author  = {M. D. Schluchter and J. D. Elashoff},
  title   = {Small-sample adjustments to tests with unbalanced repeated measures assuming several covariance structures},
  journal = {Journal of Statistical Computation and Simulation},
  year    = {1990},
  volume  = {37},
  number  = {1--2},
  pages   = {69--87},
  doi     = {10.1080/00949659008811295}
}

@manual{pnd,
  title  = {powerNormal: The power normal distribution with or without reparametrization},
  author = {K. Maruo},
  year   = {2026},
  note   = {R package version 0.1.0},
  url    = {https://github.com/kzkzmr/powerNormal}
}

@manual{bcmvr,
  title  = {bcmvr: Treatment Effect Inference Using Box-Cox Multivariate Regression},
  author = {K. Maruo},
  year   = {2026},
  note   = {R package version 0.1.0},
  url    = {https://github.com/kzkzmr/bcmvr}
}

@manual{mmrm_pkg,
  title  = {mmrm: Mixed Models for Repeated Measures},
  author = {D. {Saban{\'e}s Bov{\'e}} and L. Li and J. Dedic and D. Kelkhoff and K. Kunzmann and B. M. Lang and C. Stock and Y. Wang and D. James and J. Sidi and D. Leibovitz and D. D. Sj{\"o}berg and N. I. Krieger},
  year   = {2025},
  note   = {R package version 0.3.16},
  url    = {https://CRAN.R-project.org/package=mmrm},
  doi    = {10.32614/CRAN.package.mmrm}
}

@article{flexsurv,
  author  = {C. Jackson},
  title   = {flexsurv: A Platform for Parametric Survival Modeling in {R}},
  journal = {Journal of Statistical Software},
  year    = {2016},
  volume  = {70},
  number  = {8},
  pages   = {1--33},
  doi     = {10.18637/jss.v070.i08}
}

@article{prentice_74,
  author  = {R. L. Prentice},
  title   = {A log gamma model and its maximum likelihood estimation},
  journal = {Biometrika},
  year    = {1974},
  volume  = {61},
  number  = {3},
  pages   = {539--544},
  doi     = {10.2307/2334737}
}

@article{kenward_roger_97,
  title={Small sample inference for fixed effects from restricted maximum likelihood},
  author={Kenward, Michael G and Roger, James H},
  journal={Biometrics},
  volume={53},
  number={3},
  pages={983--997},
  year={1997},
  doi={10.2307/2533558}
}

\clearpage

\setcounter{section}{0}
\setcounter{equation}{0}
\setcounter{figure}{0}
\setcounter{table}{0}
\setcounter{footnote}{0}

\renewcommand{\thesection}{S\arabic{section}}
\renewcommand{\theequation}{S.\arabic{equation}}
\renewcommand{\thefigure}{S\arabic{figure}}
\renewcommand{\thetable}{S\arabic{table}}

\renewcommand*{\theHsection}{S\arabic{section}}
\renewcommand*{\theHequation}{S.\arabic{equation}}
\renewcommand*{\theHfigure}{S\arabic{figure}}
\renewcommand*{\theHtable}{S\arabic{table}}

\begin{center}
{\Large\bfseries
Supporting information material for\\[0.3em]
``A flexible framework for treatment effect inference in longitudinal clinical studies with skewed outcomes''\par}

\vspace{0.8em}

{\normalsize
Kazushi Maruo$^{*}$, Ryota Ishii, Yusuke Yamaguchi, \\
Toshio Shimokawa, Tomoyuki Sugimoto, and Masahiko Gosho\par}

\vspace{0.3em}

{\small $^{*}$\texttt{kazushi.maruo@gmail.com}\par}
\end{center}

\vspace{1.0em}

\section{Hessian matrix of the log-likelihood: $\mathbf{H}_g$}

For notational conventions, see Section~2.1. The Hessian matrix of the log-likelihood is
\[
\mathbf{H}_g
=
\frac{\partial^2}{\partial\bm{\theta}_g\partial\bm{\theta}_g^\top}
\,\ell_g(\bm{\theta}_g).
\]
The components of the Hessian matrix are given as follows.

\paragraph{$\lambda$--$\lambda$ block.}
\[
H_{\lambda_{gt_1}\lambda_{gt_2}}
=
-\mathbf{1}(t_1=t_2)\sum_{i_g\in\mathcal{S}_{t_1}}
\left\{z^{(2)}_{i_gt_1}\right\}_{\bm{z}_{i_g};t_1}^{\!\top}
\Sigma_{gi_g}^{-1}\bm{r}_{i_g}
-
\sum_{i_g\in\mathcal{S}_{t_1t_2}}
\left\{z^{(1)}_{i_gt_1}\right\}_{\bm{z}_{i_g};t_1}^{\!\top}
\Sigma_{gi_g}^{-1}
\left\{z^{(1)}_{i_gt_2}\right\}_{\bm{z}_{i_g};t_2},
\]
where $\mathbf{1}(\cdot)$ denotes the indicator function, which takes the value 1 if the condition inside the parentheses is satisfied and 0 otherwise, $\mathcal{S}_{\cdot}$ denotes the set of participants for whom the outcomes are observed at all time points involved in the corresponding subscript(s), and
$\{a\}_{\bm{z}_{i_g};t}$
denotes a vector of the same dimension as $\bm{z}_{i_g}$ whose component corresponding to time $t$ (i.e., the position of the observation at time $t$ in $z_{ig}$) is equal to $a$ if observed, and zero otherwise. In addition,
\begin{align*}
z^{(1)}_{i_gt}
&=
\frac{\partial z_{i_gt}}{\partial \lambda_{gt}}
=
\lambda_{gt}^{-2}
\left\{
y_{i_gt}^{\lambda_{gt}}
(\lambda_{gt}\log y_{i_gt}-1)
+1
\right\}, \\
z^{(2)}_{i_gt}
&=
\frac{\partial^2 z_{i_gt}}{\partial\lambda_{gt}^2}
=
\lambda_{gt}^{-1}y_{i_gt}^{\lambda_{gt}}(\log y_{i_gt})^2
-2\lambda_{gt}^{-2}y_{i_gt}^{\lambda_{gt}}\log y_{i_gt}
+2\lambda_{gt}^{-3}\left(y_{i_gt}^{\lambda_{gt}}-1\right).
\end{align*}

\paragraph{$\beta$--$\beta$ block.}
\[
H_{\beta_{gt_1h_1}\beta_{gt_2h_2}}
=
-\sum_{i_g\in\mathcal{S}_{t_1t_2}}
\left\{x_{i_gh_1}\right\}_{\bm{z}_{i_g};t_1}^{\!\top}
\Sigma_{gi_g}^{-1}
\left\{x_{i_gh_2}\right\}_{\bm{z}_{i_g};t_2}.
\]

\paragraph{$\alpha$--$\alpha$ block.}
\[
H_{\alpha_{gl_1}\alpha_{gl_2}}
=
-\frac{1}{2}\sum_{i_g\in\mathcal{S}_{l_1l_2}}
\mathrm{tr}\left(
A_{gi_g}^{(l_1)}\Sigma_{gi_g}^{(l_2)}
+\Sigma_{gi_g}^{-1}\Sigma_{gi_g}^{(l_1l_2)}
\right)
-\frac{1}{2}\sum_{i_g\in\mathcal{S}_{l_1l_2}}
\bm{r}_{i_g}^{\!\top}A_{gi_g}^{(l_1l_2)}\bm{r}_{i_g},
\]
where
\begin{align*}
\Sigma_g^{(l)}
&=
\frac{\partial}{\partial\alpha_{gl}}\Sigma_g,\\
\left\{\Sigma_g^{(l)}\right\}_{j,k}
&=
\begin{cases}
1, & \text{if }\alpha_{gl}\text{ corresponds to the (co)variance for time points }j\text{ and }k,\\
0, & \mathrm{otherwise},
\end{cases}
\\
A_{gi_g}^{(l)}
&=
\frac{\partial}{\partial\alpha_{gl}}\Sigma_{gi_g}^{-1}
=
-\Sigma_{gi_g}^{-1}\Sigma_{gi_g}^{(l)}\Sigma_{gi_g}^{-1},
\\
\Sigma_g^{(l_1l_2)}
&=
\bm{0}
\quad
\text{under the usual parameterization of an unstructured covariance matrix},
\\
A_{gi_g}^{(l_1l_2)}
&=
\frac{\partial^2}{\partial\alpha_{gl_1}\partial\alpha_{gl_2}}
\Sigma_{gi_g}^{-1}
\\
&=
\Sigma_{gi_g}^{-1}\left(
\Sigma_{gi_g}^{(l_1)}\Sigma_{gi_g}^{-1}\Sigma_{gi_g}^{(l_2)}
+\Sigma_{gi_g}^{(l_2)}\Sigma_{gi_g}^{-1}\Sigma_{gi_g}^{(l_1)}
-\Sigma_{gi_g}^{(l_1l_2)}
\right)\Sigma_{gi_g}^{-1}.
\end{align*}

\paragraph{Cross blocks.}
\begin{align*}
H_{\lambda_{gt_1}\beta_{gt_2h}}
&=
\sum_{i_g\in\mathcal{S}_{t_1t_2}}
\left\{x_{i_gh}\right\}_{\bm{z}_{i_g};t_2}^{\!\top}
\Sigma_{gi_g}^{-1}
\left\{z^{(1)}_{i_gt_1}\right\}_{\bm{z}_{i_g};t_1},
\\
H_{\lambda_{gt}\alpha_{gl}}
&=
-\sum_{i_g\in\mathcal{S}_{tl}}
\left\{z^{(1)}_{i_gt}\right\}_{\bm{z}_{i_g};t}^{\!\top}
A^{(l)}_{gi_g}\bm{r}_{i_g},
\\
H_{\beta_{gth}\alpha_{gl}}
&=
\sum_{i_g\in\mathcal{S}_{tl}}
\left\{x_{i_gh}\right\}_{\bm{z}_{i_g};t}^{\!\top}
A^{(l)}_{gi_g}\bm{r}_{i_g}.
\end{align*}

\section{Outer product matrix of score function: $\mathbf{J}_g$}

The matrix $\mathbf{J}_g$ is defined as
\[
\mathbf{J}_g
=
\sum_{i_g=1}^{N_g}
\left(
\frac{\partial}{\partial\bm{\theta}_g}\ell_{gi_g}(\bm{\theta}_g)
\right)
\left(
\frac{\partial}{\partial\bm{\theta}_g}\ell_{gi_g}(\bm{\theta}_g)
\right)^{\!\top},
\]
where $\ell_{gi_g}(\bm{\theta}_g)$ is the log-likelihood for the $i_g$-th participant. The components of the matrix are given as follows.

\paragraph{$\lambda$--$\lambda$ block.}
\begin{align*}
J_{\lambda_{gt_1}\lambda_{gt_2}}
&=
\sum_{i_g\in\mathcal{S}_{t_1t_2}}
(\log y_{i_gt_1})(\log y_{i_gt_2})
-\sum_{i_g\in\mathcal{S}_{t_1t_2}}
(\log y_{i_gt_1})
\left\{z^{(1)}_{i_gt_2}\right\}_{\bm{z}_{i_g};t_2}^{\!\top}
\Sigma_{gi_g}^{-1}\bm{r}_{i_g}
\\
&\quad
-\sum_{i_g\in\mathcal{S}_{t_1t_2}}
(\log y_{i_gt_2})
\left\{z^{(1)}_{i_gt_1}\right\}_{\bm{z}_{i_g};t_1}^{\!\top}
\Sigma_{gi_g}^{-1}\bm{r}_{i_g}
+\sum_{i_g\in\mathcal{S}_{t_1t_2}}
\left\{z^{(1)}_{i_gt_1}\right\}_{\bm{z}_{i_g};t_1}^{\!\top}
\Sigma_{gi_g}^{-1}\bm{r}_{i_g}\bm{r}_{i_g}^{\top}
\Sigma_{gi_g}^{-1}
\left\{z^{(1)}_{i_gt_2}\right\}_{\bm{z}_{i_g};t_2}.
\end{align*}

\paragraph{$\beta$--$\beta$ block.}
\[
J_{\beta_{gt_1h_1}\beta_{gt_2h_2}}
=
\sum_{i_g\in\mathcal{S}_{t_1t_2}}
\left\{x_{i_gh_1}\right\}_{\bm{z}_{i_g};t_1}^{\!\top}
\Sigma_{gi_g}^{-1}\bm{r}_{i_g}\bm{r}_{i_g}^{\top}
\Sigma_{gi_g}^{-1}
\left\{x_{i_gh_2}\right\}_{\bm{z}_{i_g};t_2}.
\]

\paragraph{$\alpha$--$\alpha$ block.}
\begin{align*}
J_{\alpha_{gl_1}\alpha_{gl_2}}
&=
\frac{1}{4}\sum_{i_g\in\mathcal{S}_{l_1l_2}}
\mathrm{tr}\left(\Sigma_{gi_g}^{-1}\Sigma_{gi_g}^{(l_1)}\right)\,
\mathrm{tr}\left(\Sigma_{gi_g}^{-1}\Sigma_{gi_g}^{(l_2)}\right)
+\frac{1}{4}\sum_{i_g\in\mathcal{S}_{l_1l_2}}
\mathrm{tr}\left(\Sigma_{gi_g}^{-1}\Sigma_{gi_g}^{(l_1)}\right)\,
\bm{r}_{i_g}^{\top}A_{gi_g}^{(l_2)}\bm{r}_{i_g}
\\
&\quad
+\frac{1}{4}\sum_{i_g\in\mathcal{S}_{l_1l_2}}
\mathrm{tr}\left(\Sigma_{gi_g}^{-1}\Sigma_{gi_g}^{(l_2)}\right)\,
\bm{r}_{i_g}^{\top}A_{gi_g}^{(l_1)}\bm{r}_{i_g}
+\frac{1}{4}\sum_{i_g\in\mathcal{S}_{l_1l_2}}
\bm{r}_{i_g}^{\top}A_{gi_g}^{(l_1)}\bm{r}_{i_g}\,
\bm{r}_{i_g}^{\top}A_{gi_g}^{(l_2)}\bm{r}_{i_g}.
\end{align*}

\paragraph{Cross blocks.}
\begin{align*}
J_{\lambda_{gt_1}\beta_{gt_2h}}
&=
\sum_{i_g\in\mathcal{S}_{t_1t_2}}
(\log y_{i_gt_1})
\left\{x_{i_gh}\right\}_{\bm{z}_{i_g};t_2}^{\!\top}
\Sigma_{gi_g}^{-1}\bm{r}_{i_g}
-
\sum_{i_g\in\mathcal{S}_{t_1t_2}}
\left\{x_{i_gh}\right\}_{\bm{z}_{i_g};t_2}^{\!\top}
\Sigma_{gi_g}^{-1}\bm{r}_{i_g}\bm{r}_{i_g}^{\top}
\Sigma_{gi_g}^{-1}
\left\{z^{(1)}_{i_gt_1}\right\}_{\bm{z}_{i_g};t_1},
\\
J_{\lambda_{gt}\alpha_{gl}}
&=
-\frac{1}{2}\sum_{i_g\in\mathcal{S}_{tl}}
(\log y_{i_gt})\,\mathrm{tr}\left(\Sigma_{gi_g}^{-1}\Sigma_{gi_g}^{(l)}\right)
-\frac{1}{2}\sum_{i_g\in\mathcal{S}_{tl}}
(\log y_{i_gt})\,\bm{r}_{i_g}^{\top}A_{gi_g}^{(l)}\bm{r}_{i_g}
\\
&\quad
+\frac{1}{2}\sum_{i_g\in\mathcal{S}_{tl}}
\left\{z^{(1)}_{i_gt}\right\}_{\bm{z}_{i_g};t}^{\!\top}
\Sigma_{gi_g}^{-1}\bm{r}_{i_g}\,
\mathrm{tr}\left(\Sigma_{gi_g}^{-1}\Sigma_{gi_g}^{(l)}\right)
+\frac{1}{2}\sum_{i_g\in\mathcal{S}_{tl}}
\left\{z^{(1)}_{i_gt}\right\}_{\bm{z}_{i_g};t}^{\!\top}
\Sigma_{gi_g}^{-1}\bm{r}_{i_g}\,
\bm{r}_{i_g}^{\top}A_{gi_g}^{(l)}\bm{r}_{i_g},
\\
J_{\beta_{gth}\alpha_{gl}}
&=
-\frac{1}{2}\sum_{i_g\in\mathcal{S}_{tl}}
\left\{x_{i_gh}\right\}_{\bm{z}_{i_g};t}^{\!\top}
\Sigma_{gi_g}^{-1}\bm{r}_{i_g}\,
\mathrm{tr}\left(\Sigma_{gi_g}^{-1}\Sigma_{gi_g}^{(l)}\right)
-\frac{1}{2}\sum_{i_g\in\mathcal{S}_{tl}}
\left\{x_{i_gh}\right\}_{\bm{z}_{i_g};t}^{\!\top}
\Sigma_{gi_g}^{-1}\bm{r}_{i_g}\,
\bm{r}_{i_g}^{\top}A_{gi_g}^{(l)}\bm{r}_{i_g}.
\end{align*}

\section{Gradient vector for median function}

The gradient vector for the median function, $\xi_{gt}(\bm{\theta}_g)$, is given by
\[
\nabla_{\bm{\theta}_g}\xi_{gt}(\bm{\theta}_g)
=
\frac{\partial}{\partial\bm{\theta}_g}\xi_{gt}(\bm{\theta}_g)
=
\left(
\begin{matrix}
\lambda_{gt}^{-2}\xi_{gt}
\left(1-\lambda_{gt}\log\xi_{gt}-\xi_{gt}^{-\lambda_{gt}}\right)\bm{e}_t \\
\xi_{gt}^{1-\lambda_{gt}}\bm{e}_t \\
\bar{x}_1\,\xi_{gt}^{1-\lambda_{gt}}\bm{e}_t \\
\vdots \\
\bar{x}_K\,\xi_{gt}^{1-\lambda_{gt}}\bm{e}_t \\
\bm{0}_M
\end{matrix}
\right),
\]
where $\bm{e}_t$ denotes the $t$th canonical basis vector in $\mathbb{R}^T$, that is, a vector with 1 in the $t$th position and 0 elsewhere, and $\mathbf{0}_M$ denotes a zero vector of length $M$.

\section{Gradient vector for the logit-transformed probability-based effect measure}

The gradient vector for the logit-transformed probability-based effect measure is given by
\begin{align*}
\nabla_{\bm{\theta}_{g_1 g_2}}
\eta^{(p)}_{g_1 g_2 t}
(\bm{\theta}_{g_1 g_2})
&=
\frac{\partial }
{\partial \bm{\theta}_{g_1 g_2}}\eta^{(p)}_{g_1 g_2 t}(\bm{\theta}_{g_1 g_2})
=
\frac{\partial}
{\partial \bm{\theta}_{g_1 g_2}}
\log\!\left\{
\frac{\Delta^{(p)}_{g_1 g_2 t}}
{1-\Delta^{(p)}_{g_1 g_2 t}}
\right\}
\\
&=
\left(
\frac{1}{\Delta^{(p)}_{g_1 g_2 t}}
+\frac{1}{1-\Delta^{(p)}_{g_1 g_2 t}}
\right)
\frac{\partial }
{\partial \bm{\theta}_{g_1 g_2}}
\Delta^{(p)}_{g_1 g_2 t}(\bm{\theta}_{g_1 g_2}).
\end{align*}
The components are given as follows.
\begin{align*}
\frac{\partial}{\partial \lambda_{g_1 t}}
\Delta^{(p)}_{g_1 g_2 t}(\bm{\theta}_{g_1 g_2})
&=
\int_{0}^{\infty}
y^{\lambda_{g_2 t}-1}
z_{g_1 t}
f_{\mathrm{N}}(z_{g_1 t};\mu_{g_1 t},\sigma_{g_1 t}^2)
f_{\mathrm{N}}(z_{g_2 t};\mu_{g_2 t},\sigma_{g_2 t}^2)
\,dy,
\\[4pt]
\frac{\partial}{\partial \lambda_{g_2 t}}
\Delta^{(p)}_{g_1 g_2 t}(\bm{\theta}_{g_1 g_2})
&=
\int_{0}^{\infty}
y^{\lambda_{g_2 t}-1}
\log y
\frac{z_{g_2 t}-\mu_{g_2 t}}{\sigma_{g_2 t}}
F_{\mathrm{N}}(z_{g_1 t};\mu_{g_1 t},\sigma_{g_1 t}^2)
f_{\mathrm{N}}(z_{g_2 t};\mu_{g_2 t},\sigma_{g_2 t}^2)
\,dy,
\\[4pt]
\frac{\partial}{\partial \beta_{g_1 t k}}
\Delta^{(p)}_{g_1 g_2 t}(\bm{\theta}_{g_1 g_2})
&=
-\bar{x}_k
\int_{0}^{\infty}
y^{\lambda_{g_2 t}-1}
f_{\mathrm{N}}(z_{g_1 t};\mu_{g_1 t},\sigma_{g_1 t}^2)
f_{\mathrm{N}}(z_{g_2 t};\mu_{g_2 t},\sigma_{g_2 t}^2)
\,dy,
\\[4pt]
\frac{\partial}{\partial \beta_{g_2 t k}}
\Delta^{(p)}_{g_1 g_2 t}(\bm{\theta}_{g_1 g_2})
&=
\frac{\bar{x}_k}{\sigma_{g_2 t}}
\int_{0}^{\infty}
\frac{z_{g_2 t}-\mu_{g_2 t}}{\sigma_{g_2 t}}
y^{\lambda_{g_2 t}-1}
F_{\mathrm{N}}(z_{g_1 t};\mu_{g_1 t},\sigma_{g_1 t}^2)
f_{\mathrm{N}}(z_{g_2 t};\mu_{g_2 t},\sigma_{g_2 t}^2)
\,dy,
\\[4pt]
\frac{\partial}{\partial \sigma_{g_1 t}^2}
\Delta^{(p)}_{g_1 g_2 t}(\bm{\theta}_{g_1 g_2})
&=
-\frac{1}{2\sigma_{g_1 t}}
\int_{0}^{\infty}
\frac{z_{g_1 t}-\mu_{g_1 t}}{\sigma_{g_1 t}}
y^{\lambda_{g_2 t}-1}
f_{\mathrm{N}}(z_{g_1 t};\mu_{g_1 t},\sigma_{g_1 t}^2)
f_{\mathrm{N}}(z_{g_2 t};\mu_{g_2 t},\sigma_{g_2 t}^2)
\,dy,
\\[4pt]
\frac{\partial}{\partial \sigma_{g_2 t}^2}
\Delta^{(p)}_{g_1 g_2 t}(\bm{\theta}_{g_1 g_2})
&=
\frac{1}{2\sigma_{g_2 t}^2}
\int_{0}^{\infty}
\frac{(z_{g_2 t}-\mu_{g_2 t})^2}{\sigma_{g_2 t}^2}
y^{\lambda_{g_2 t}-1}
F_{\mathrm{N}}(z_{g_1 t};\mu_{g_1 t},\sigma_{g_1 t}^2)
f_{\mathrm{N}}(z_{g_2 t};\mu_{g_2 t},\sigma_{g_2 t}^2)
\,dy
\\
&\quad-
\Delta^{(p)}_{g_1 g_2 t}(\bm{\theta}_{g_1 g_2}).
\end{align*}

These components are evaluated using numerical integration.

\clearpage
\section{Additional simulation results under 30\% dropout}
\subsection*{Bias of treatment effect under null hypothesis}
\begin{figure*}[h]
\centerline{ \includegraphics{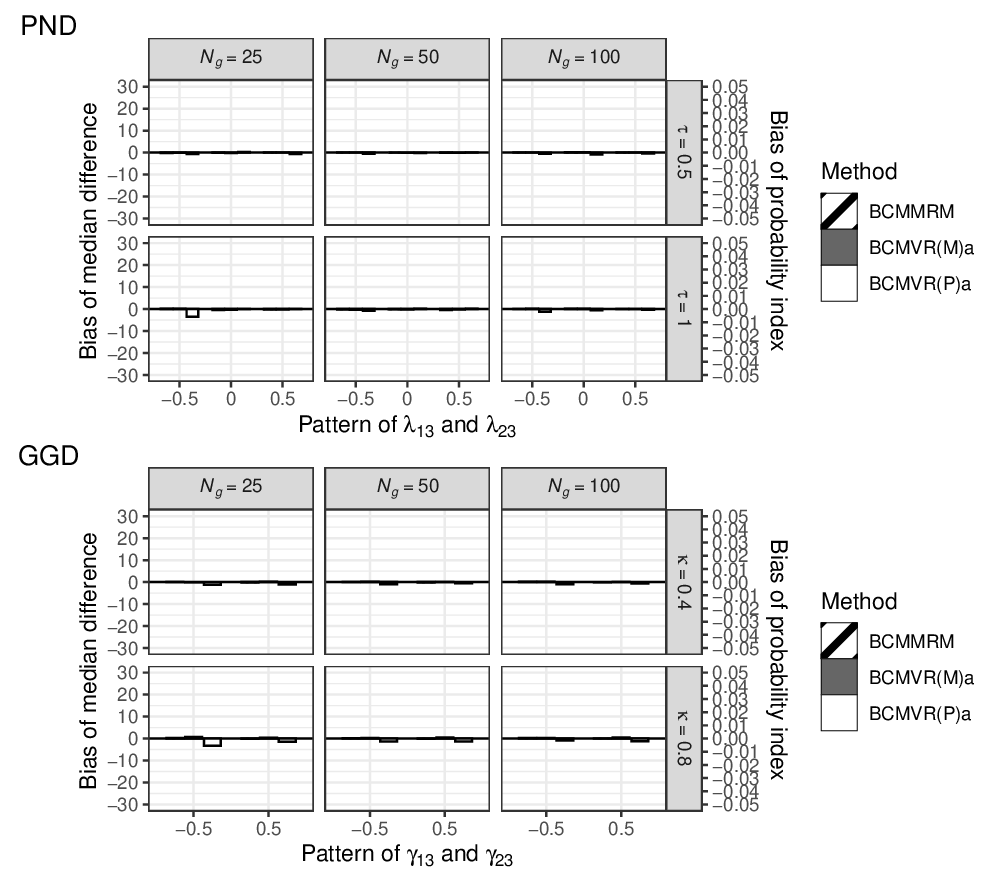}}
\caption{Empirical bias of treatment effect under the null hypothesis.
Biases of the median difference for BCMMRM and BCMVR(M)a are shown on the left vertical axis, while biases of the probability-based measure for BCMVR(P)a are shown on the right vertical axis.}
  \label{fig:bias_n}
\end{figure*}

\clearpage
\subsection*{Standard error ratio}

\begin{figure*}[h!]
\centerline{ \includegraphics{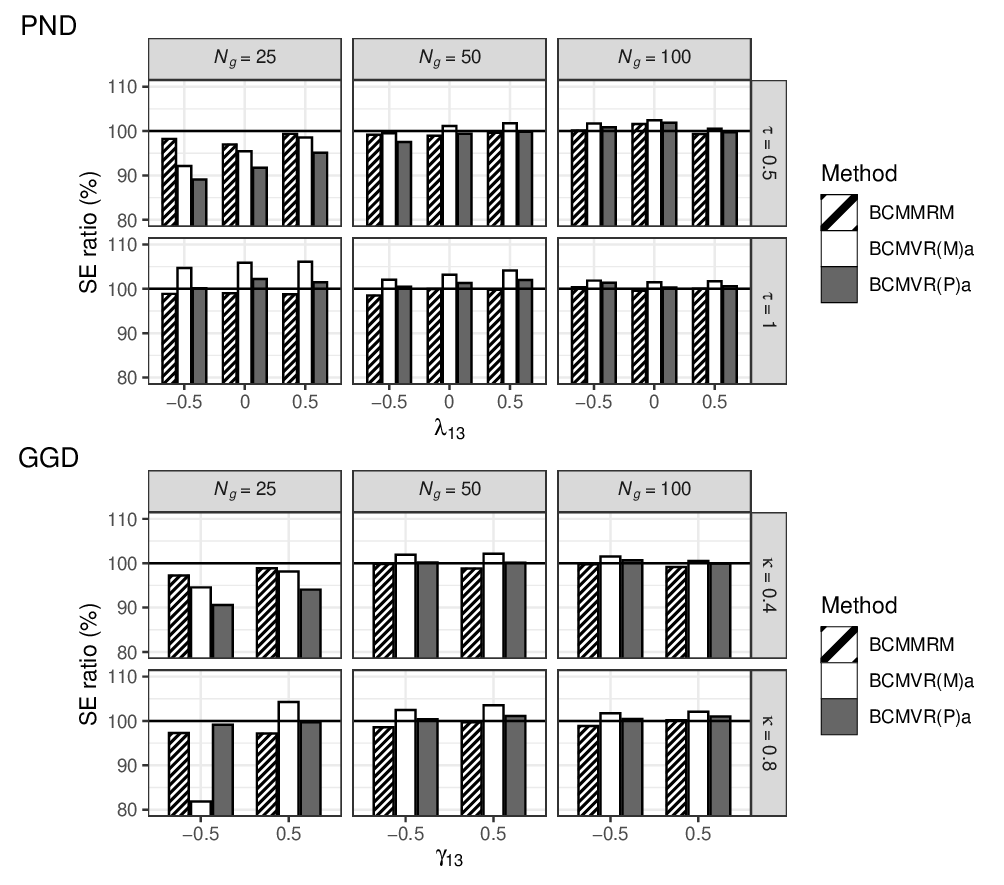}}
\caption{Standard error ratio (\{average of estimated standard errors\}/\{SD of estimated treatment effect\}$\times 100$) under the null hypothesis.}
  \label{fig:ser_n}
\end{figure*}

\begin{figure*}[h]
\centerline{ \includegraphics{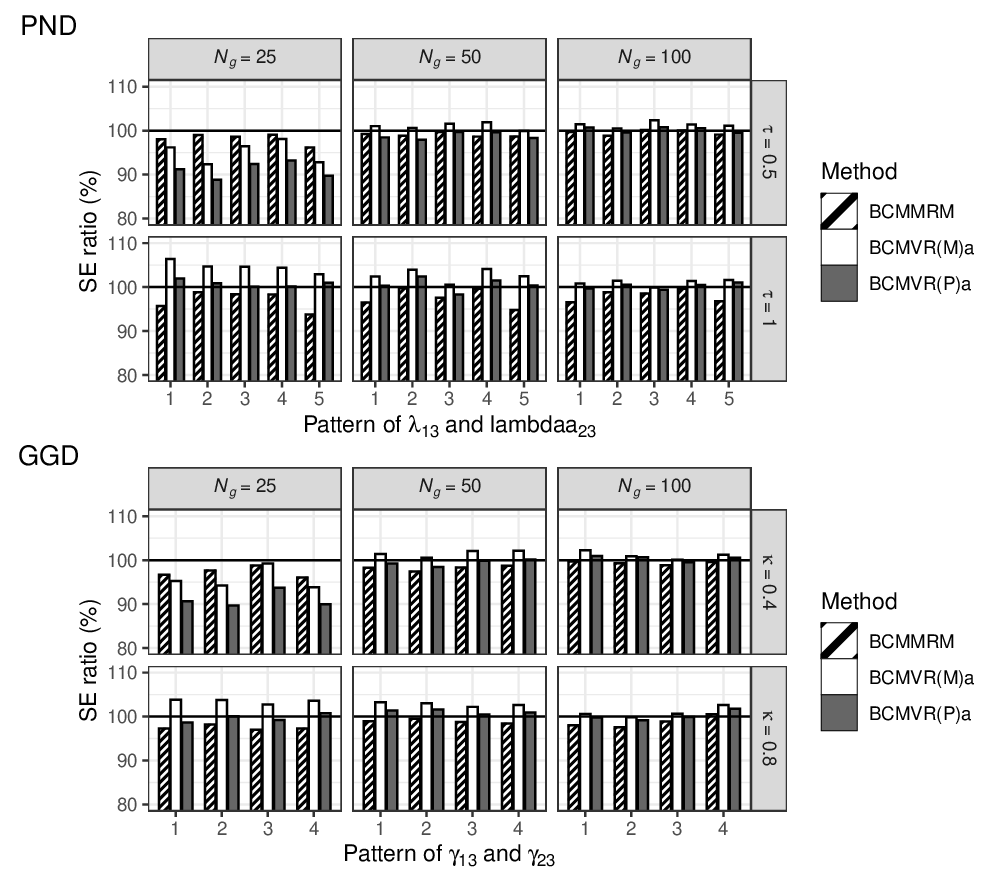}}
\caption{Standard error ratio (\{average of estimated standard errors\}/\{SD of estimated treatment effect\}$\times 100$)  under the alternative hypothesis.}
  \label{fig:ser_a}
\end{figure*}

\clearpage
\subsection*{Empirical coverage probability}
\begin{figure*}[h]
\centerline{ \includegraphics{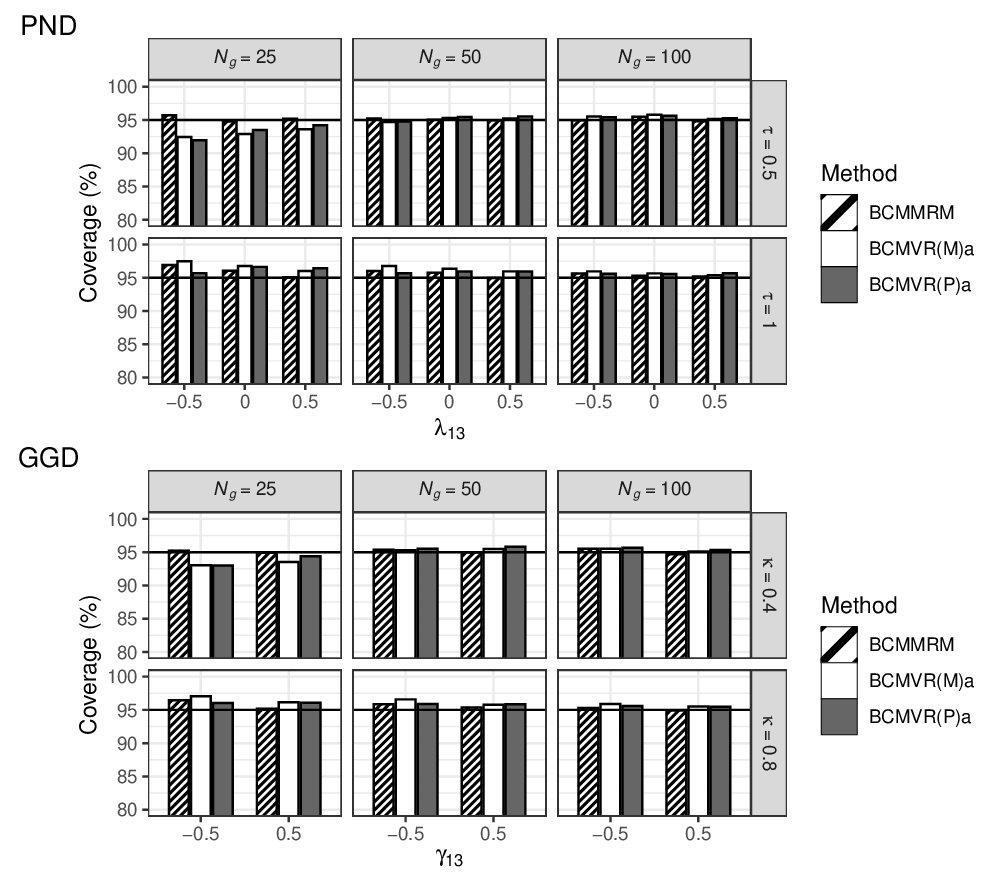}}
\caption{Empirical coverage probability of treatment effect under the null hypothesis.}
  \label{fig:cp_n}
\end{figure*}

\begin{figure*}[h]
\centerline{ \includegraphics{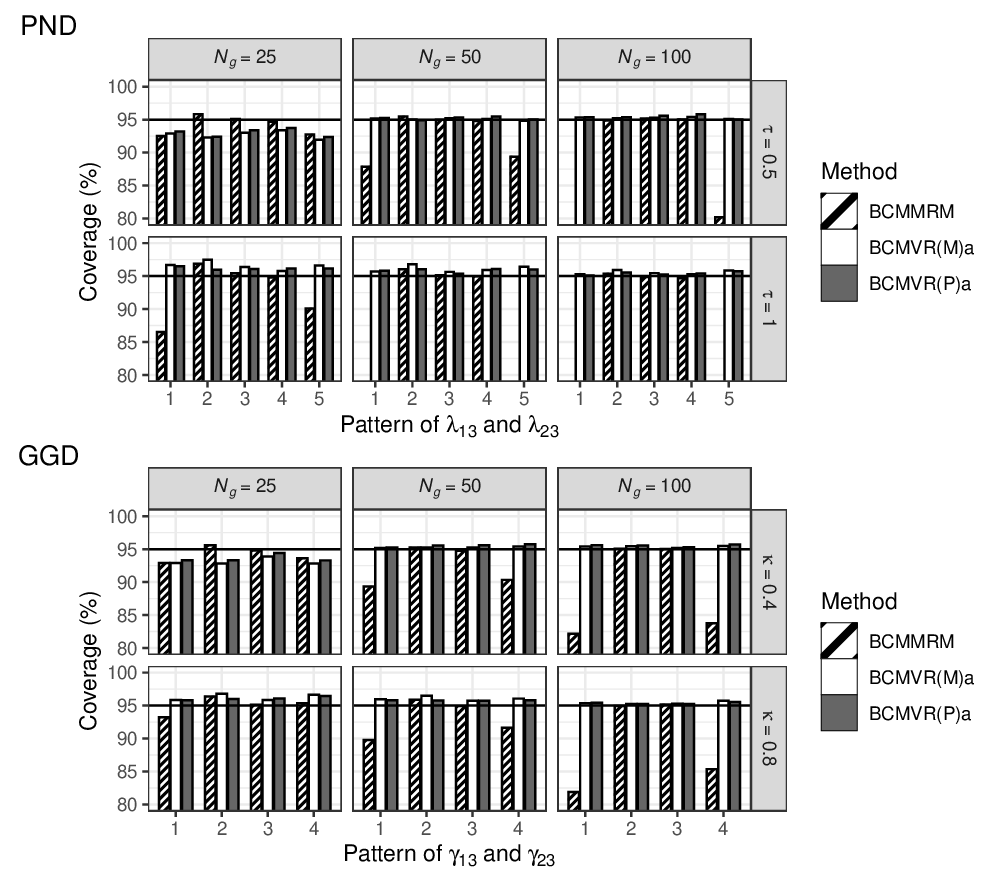}}
\caption{Empirical coverage probability of treatment effect under the alternative hypothesis.}
  \label{fig:cp_a}
\end{figure*}
\clearpage

\section{Complete simulation results under no dropout}
\subsection*{Model selection between BCMMRM and BCMVR}
\begin{table*}[h!]
\caption{Proportion of simulations in which the BCMVR model was selected over the BCMMRM model based on the likelihood ratio test.}
\centering
\setlength{\tabcolsep}{4pt}

\begin{tabular}{
lcc
S[table-format=3.1]
S[table-format=2.1]
S[table-format=3.1]
S[table-format=3.1]
S[table-format=2.1]
S[table-format=2.1]
S[table-format=2.1]
S[table-format=3.1]
}
\hline

Dist. & Scale & $n$ &
\multicolumn{3}{c}{Null scenarios} &
\multicolumn{5}{c}{Alternative scenarios} \\

\cline{4-6} \cline{7-11}

& & & {$-0.5$} & {$0$} & {$0.5$} &
{1} & {2} & {3} & {4} & {5} \\

\hline

PND & 0.5 & 25  & 11.3 &  9.0 & 11.0 &  25.5 & 11.4 &  8.7 & 11.8 &  23.1 \\
PND & 0.5 & 50  & 11.8 &  6.6 & 12.6 &  44.4 & 13.1 &  7.2 & 13.2 &  42.5 \\
PND & 0.5 & 100 & 16.0 &  5.8 & 17.4 &  80.4 & 19.6 &  5.7 & 20.9 &  79.4 \\

PND & 1.0 & 25  & 18.5 &  9.9 & 18.9 &  74.5 & 19.8 & 10.2 & 20.0 &  72.6 \\
PND & 1.0 & 50  & 27.4 &  7.0 & 27.4 &  97.6 & 29.3 &  6.9 & 30.2 &  97.4 \\
PND & 1.0 & 100 & 53.7 &  6.1 & 56.6 & 100.0 & 57.9 &  5.5 & 59.4 & 100.0 \\

\hline

GGD & 0.4 & 25  & 10.5 & \multicolumn{1}{c}{--} & 10.3 &
20.6 & 11.0 & 11.7 & 20.6 & \multicolumn{1}{c}{--} \\

GGD & 0.4 & 50  &  9.7 & \multicolumn{1}{c}{--} &  9.8 &
35.1 & 10.3 & 10.6 & 33.2 & \multicolumn{1}{c}{--} \\

GGD & 0.4 & 100 & 12.8 & \multicolumn{1}{c}{--} & 12.4 &
68.3 & 13.8 & 13.9 & 66.2 & \multicolumn{1}{c}{--} \\

GGD & 0.8 & 25  & 11.6 & \multicolumn{1}{c}{--} & 11.3 &
20.9 & 11.2 & 11.4 & 19.8 & \multicolumn{1}{c}{--} \\

GGD & 0.8 & 50  &  9.8 & \multicolumn{1}{c}{--} & 10.4 &
34.5 & 10.1 & 10.1 & 33.8 & \multicolumn{1}{c}{--} \\

GGD & 0.8 & 100 & 12.8 & \multicolumn{1}{c}{--} & 13.1 &
67.3 & 13.5 & 12.8 & 67.5 & \multicolumn{1}{c}{--} \\

\hline
\end{tabular}
\end{table*}

\subsection*{Convergence rates}
\begin{table}[h!]
\caption{Convergence rates (\%) of the estimation algorithm across simulation scenarios.}
\centering
\label{tab:conv_nm}
\begin{tabular}{llrrrrr}
\toprule
Dist. & Scenario & Min. & Q1 & Median & Q3 & Max. \\
\midrule
PND & $N_g=25$, $\tau=0.5$ & 95.2 & 96.0 & 96.6 & 96.7 & 97.1 \\
PND & Other                & 99.2 & 100.0 & 100.0 & 100.0 & 100.0 \\
GGD & $N_g=25$, $\kappa=0.4$ & 98.2 & 98.2 & 98.2 & 98.2 & 98.4 \\
GGD & Other                & 99.7 & 100.0 & 100.0 & 100.0 & 100.0 \\
\bottomrule
\end{tabular}
\end{table}

\clearpage

\subsection*{Empirical type I error}
\begin{figure*}[h!]
\centerline{ \includegraphics{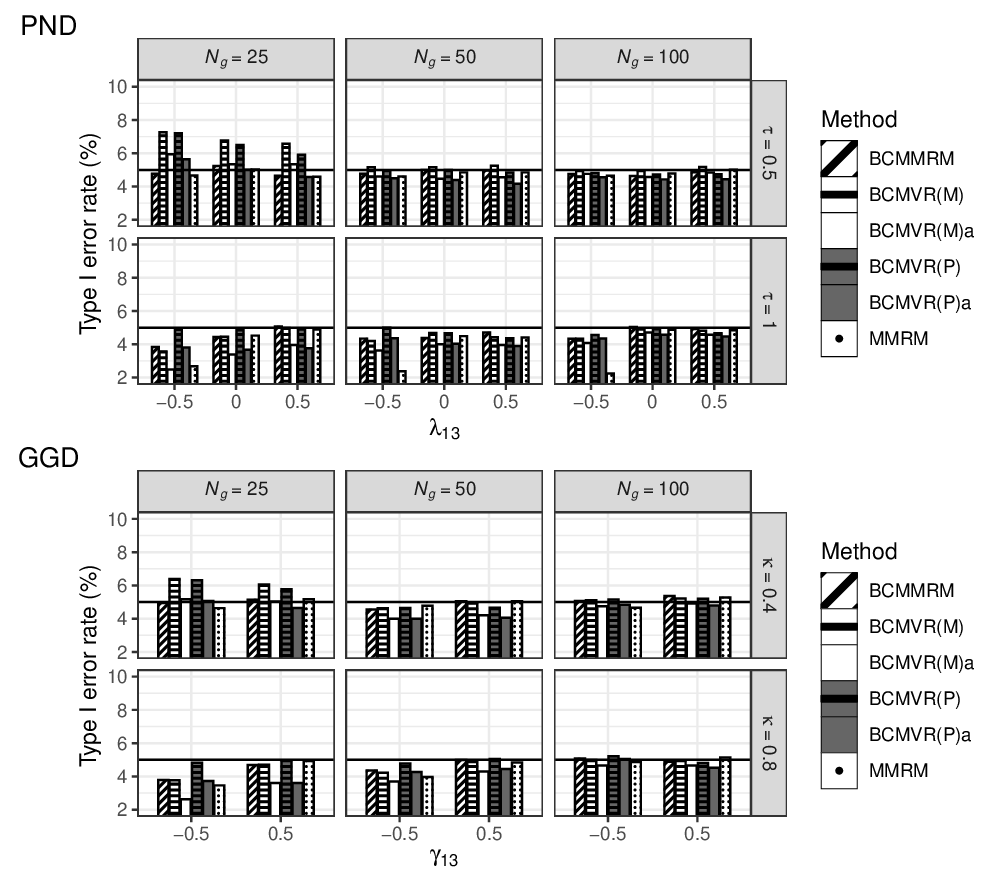}}
\caption{Empirical type I error.}
  \label{fig:test_size_nm}
\end{figure*}

\clearpage
\subsection*{Empirical power}
\begin{figure*}[h!]
\centerline{ \includegraphics{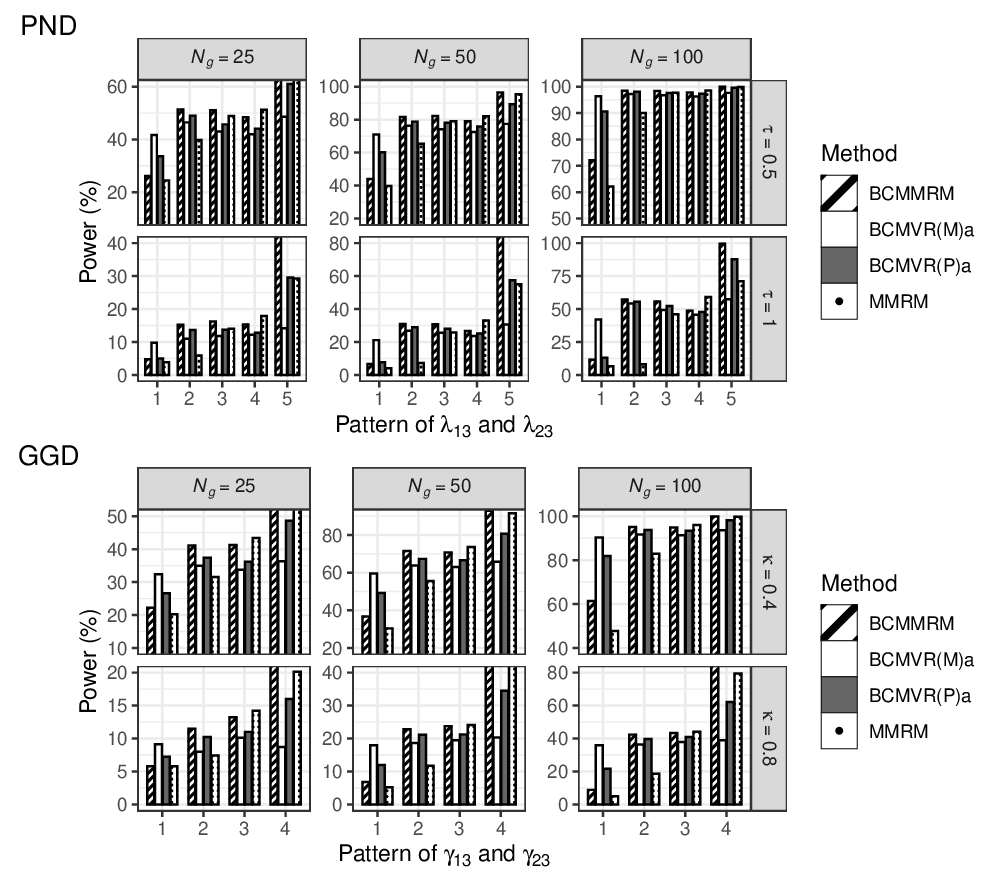}}
\caption{Empirical power.}
  \label{fig:power_nm}
\end{figure*}

\clearpage
\subsection*{Bias of treatment effect}
\begin{figure*}[h]
\centerline{ \includegraphics{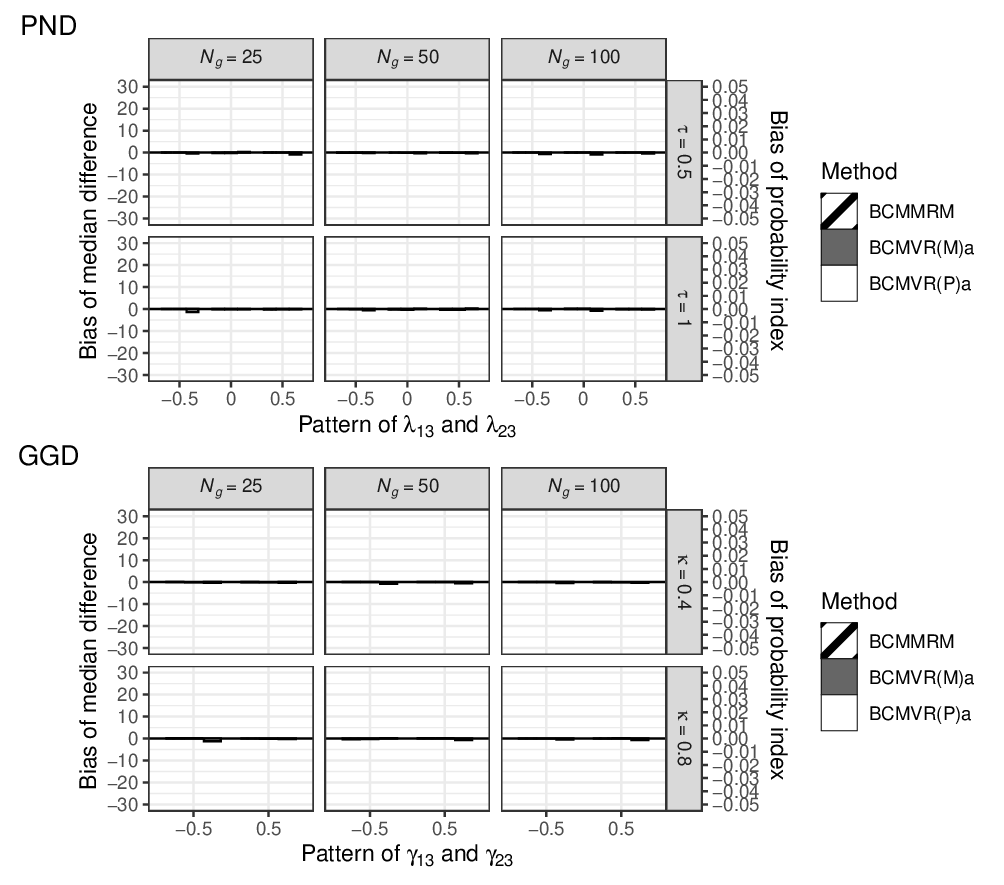}}
\caption{Empirical bias of treatment effect under the null hypothesis.
Biases of the median difference for BCMMRM and BCMVR(M)a are shown on the left vertical axis, while biases of the probability-based measure for BCMVR(P)a are shown on the right vertical axis.}
  \label{fig:bias_n_nm}
\end{figure*}
\begin{figure*}[h]
\centerline{ \includegraphics{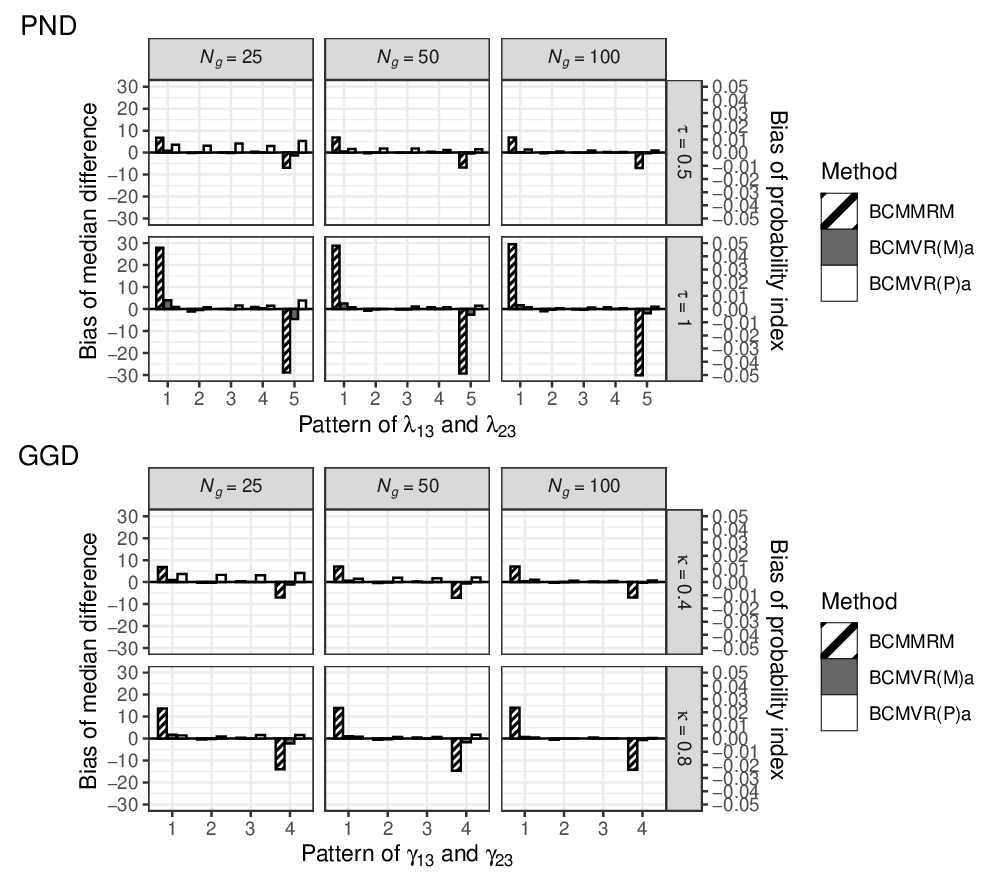}}
\caption{Empirical bias of treatment effect under the alternative hypothesis.
Biases of the median difference for BCMMRM and BCMVR(M)a are shown on the left vertical axis, while biases of the probability-based measure for BCMVR(P)a are shown on the right vertical axis.}
  \label{fig:bias_a_nm}
\end{figure*}

\clearpage
\subsection*{Standard error ratio}

\begin{figure*}[h!]
\centerline{ \includegraphics{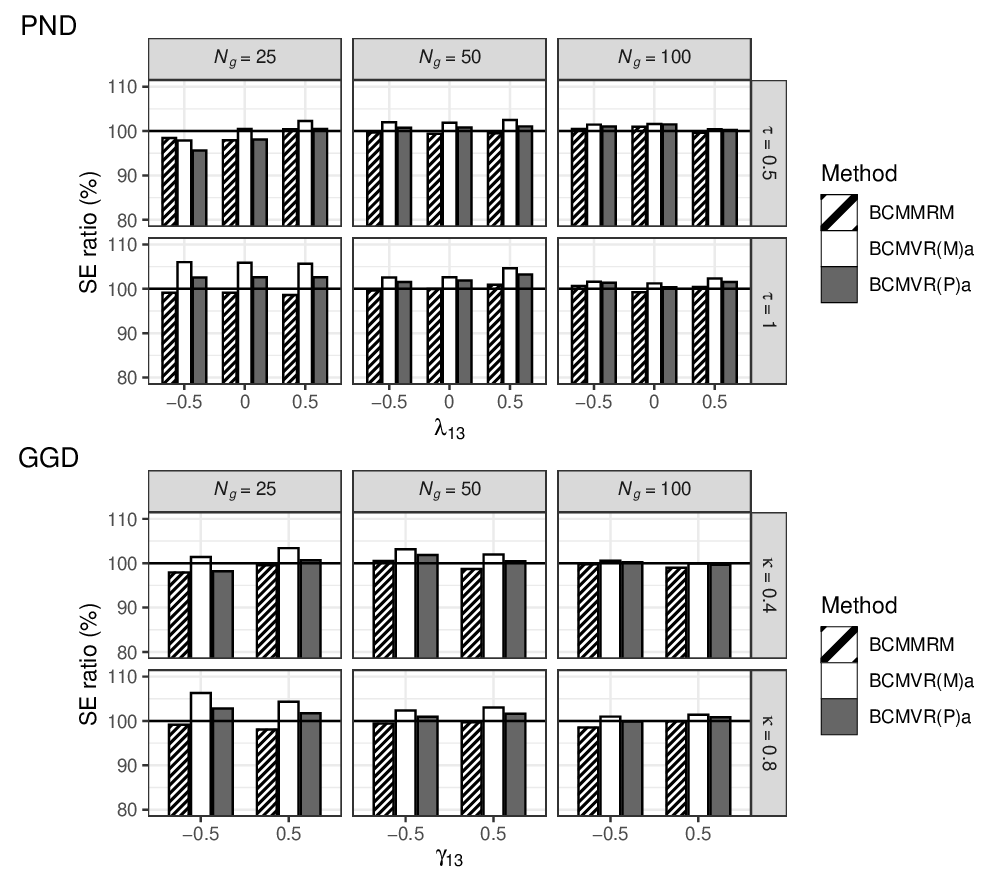}}
\caption{Standard error ratio (\{average of estimated standard errors\}/\{SD of estimated treatment effect\}$\times 100$) under the null hypothesis.}
  \label{fig:ser_n_nm}
\end{figure*}

\begin{figure*}[h]
\centerline{ \includegraphics{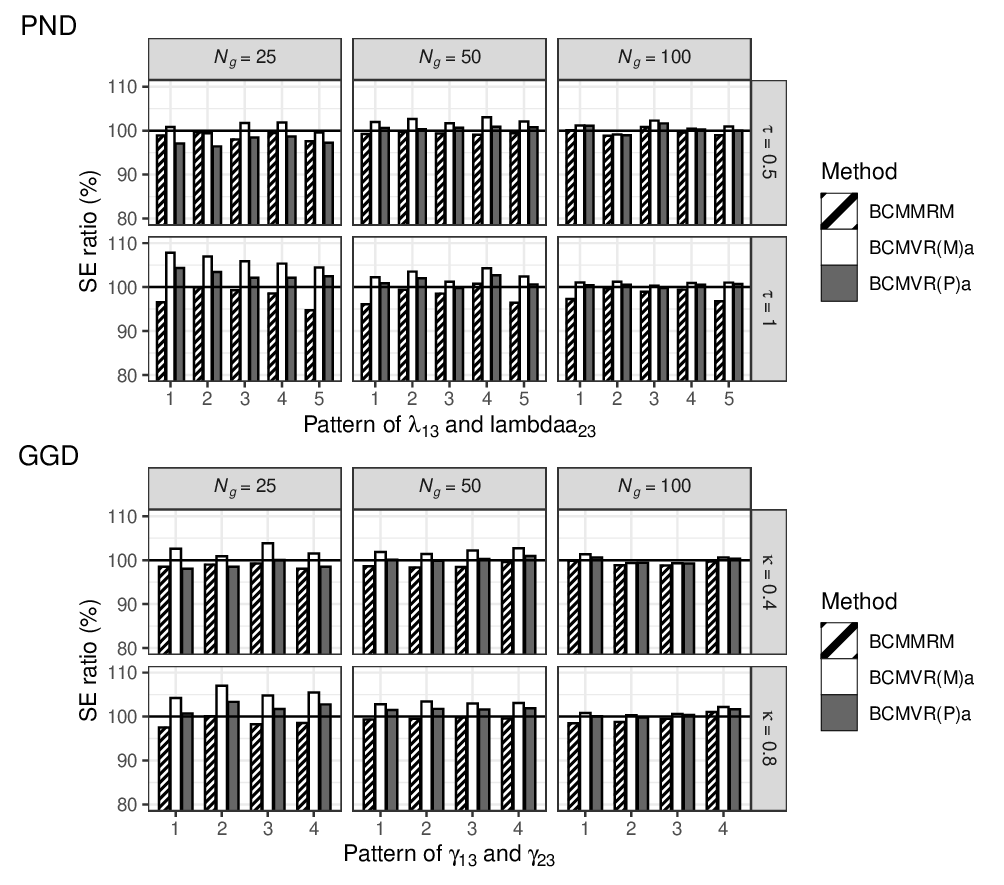}}
\caption{Standard error ratio (\{average of estimated standard errors\}/\{SD of estimated treatment effect\}$\times 100$)  under the alternative hypothesis.}
  \label{fig:ser_a_nm}
\end{figure*}

\clearpage
\subsection*{Empirical coverage probability}
\begin{figure*}[h]
\centerline{ \includegraphics{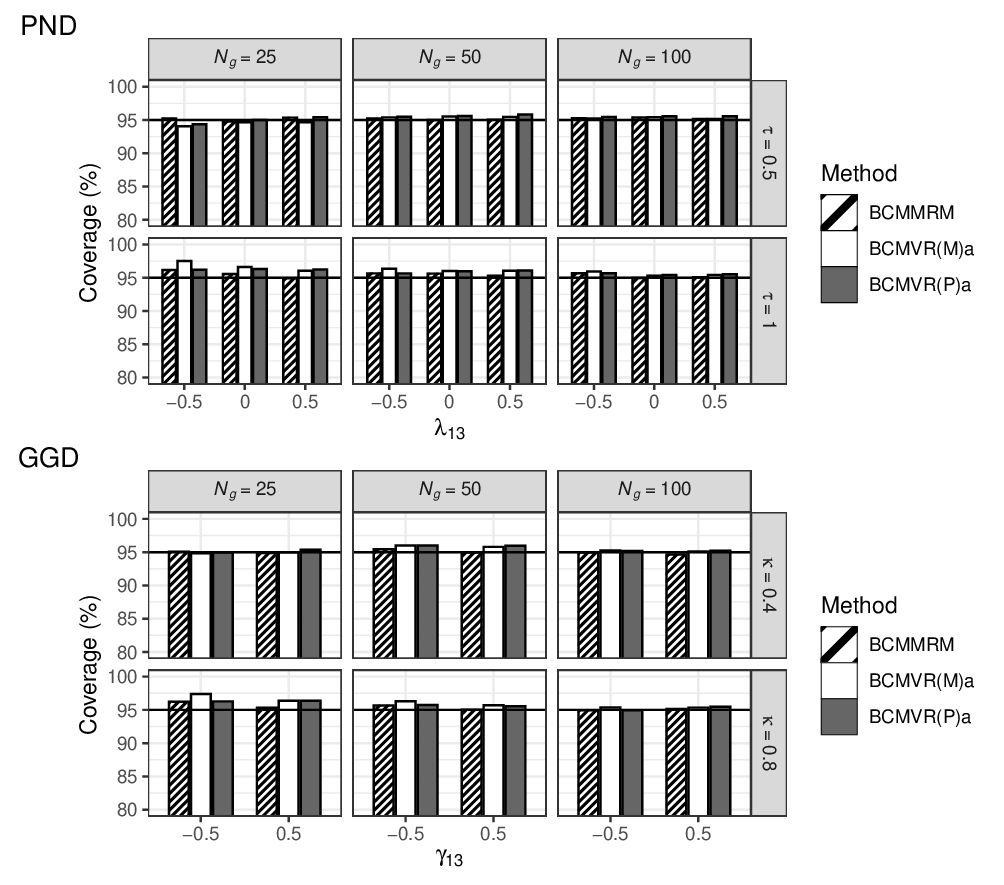}}
\caption{Empirical coverage probability of treatment effect under the null hypothesis.}
  \label{fig:cp_n_nm}
\end{figure*}

\begin{figure*}[h]
\centerline{ \includegraphics{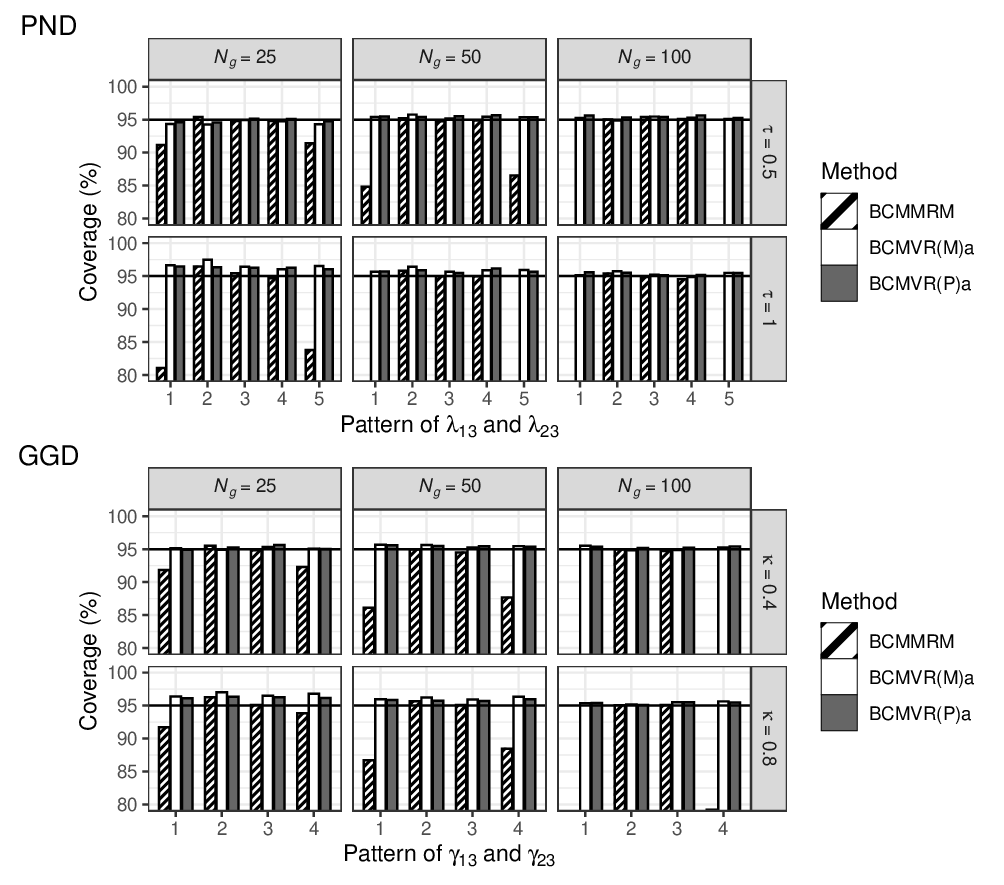}}
\caption{Empirical coverage probability of treatment effect under the alternative hypothesis.}
  \label{fig:cp_a_nm}
\end{figure*}

\end{document}